\documentclass[10pt,journal,final]{IEEEtran}
\usepackage{amsmath,amsfonts}
\usepackage{amsthm}
\usepackage{amssymb}
\usepackage{algorithmic}
\usepackage{algorithm}
\usepackage{array}
\usepackage{mathrsfs}
\usepackage[caption=false,font=footnotesize,labelfont=rm,textfont=rm]{subfig}
\usepackage{textcomp}
\usepackage{stfloats}
\usepackage{url}
\usepackage{mathtools}
\usepackage{verbatim}
\usepackage{graphicx}
\usepackage{cite}
\usepackage{autobreak}
\usepackage{nicematrix}
\usepackage{arydshln}
\usepackage{latexsym}
\usepackage{booktabs}
\usepackage{subeqnarray}
\usepackage{cases}
\usepackage{nomencl}
\usepackage{graphicx}
\makenomenclature
\allowdisplaybreaks[4]

\usepackage{etoolbox}
\renewcommand\nomgroup[1]{%
	\item[\bfseries
	\ifstrequal{#1}{A}{Abbreviations}{%
	\ifstrequal{#1}{B}{Indices and Sets}{%
	\ifstrequal{#1}{C}{Parameters}{%
	\ifstrequal{#1}{D}{Variables}{%
	\ifstrequal{#1}{E}{Functions}{}}}}}%
]}

\begin{document}
\title{Byzantine-Resilient Distributed P2P Energy Trading via Spatial-Temporal Anomaly Detection}

\author{Junhong~Liu,~\IEEEmembership{Member,~IEEE},~Qinfei~Long,~\IEEEmembership{Member,~IEEE},\\~Rong-Peng~Liu,~\IEEEmembership{Member,~IEEE},~Wenjie~Liu,~\IEEEmembership{Member,~IEEE},~and~Yunhe~Hou,~\IEEEmembership{Senior~Member,~IEEE}
\thanks{This work was supported in part by the Shenzhen-Hong Kong-Macau Science and Technology Program (Type C) under Grant SDGX202205303000214, in part by the National Natural Science Foundation of China (NSFC) under Grant 52437005, in part by the Research Grants Council of Hong Kong under Grant GRF 17201524, and in part by the Fonds de recherche du Québec-secteur Nature et technologies (FRQ-Secteur NT) under Grant FRQ-Secteur NT 367013 and Grant FRQ-Secteur NT 334636. \textit{(Corresponding author: Yunhe Hou.)}}
        
\thanks{Junhong Liu, Qinfei Long, and Yunhe Hou are with the Department of Electrical and Electronic Engineering, The University of Hong Kong, Hong Kong SAR, China (e-mail: {jhliu, qflong, yhhou}@eee.hku.hk).}
\thanks{Rong-Peng Liu is with the Department of Electrical and Computer Engineering, McGill University, Montreal, QC H3A 0E9, Canada (e-mail: rpliu@eee.hku.hk).}
\thanks{Wenjie Liu is with the School of Mechanical and Electrical Engineering, Guangzhou University, Guangzhou, China (e-mail: lwj1993@gzhu.edu.cn).}
\vspace{-2.0em}}

\markboth{}%
{Shell \MakeLowercase{\textit{et al.}}: A Sample Article Using IEEEtran.cls for IEEE Journals}
\maketitle
\begin{abstract} 
Distributed peer-to-peer (P2P) energy trading mandates an escalating coupling between the physical power network and communication network, necessitating high-frequency sharing of real-time data among prosumers. However, this data-sharing scheme renders the system vulnerable to various malicious behaviors, as Byzantine agents can initiate cyber-attacks by injecting sophisticated false data. To better investigate the impacts of malicious Byzantine faults, this paper develops a fully distributed P2P energy trading model by accounting for the high-fidelity physical network constraints. To further detect Byzantine faults and mitigate their impacts on distributed P2P energy trading problem, we propose an online spatial-temporal anomaly detection approach by leveraging the tensor learning method, which is informed by the domain knowledge to enable awesome detection performance. Moreover, to enhance its computational efficiency, we further develop closed-form solutions for the proposed detection approach. Subsequently, we derive theoretical conditions for guaranteeing optimality and convergence of the distributed P2P energy trading problem with anomaly detection mechanisms. Results from numerical simulations validate the effectiveness, optimality, and scalability of the proposed approach.
\end{abstract}


\section{Introduction}
\IEEEPARstart{I}{n} pursuit of net-zero emissions, distributed energy resources (DER), including wind, photovoltaic, and biomass generators \cite{zhai2025advancing}, are being massively integrated into power systems. The rapid integration of DER catalyzes the emergence of localized energy solutions. DER owners can actively engage in the distributed peer-to-peer (P2P) energy trading as prosumers. Numerous benefits have been identified for this novel trading mechanism, including increased social welfare, boosted system flexibility, and guaranteed privacy preservation \cite{morstyn2018using}. Physically, the distributed P2P energy trading scheme necessitates the power system to rely more on the communication networks to facilitate seamless electricity flows among prosumers \cite{zhou2020state}. However, this reliance inevitably entails increased frequency in transmitting real-time data among prosumers \cite{chen2022blockchain,liu2023privacy}. 

Ideally, where all prosumers in the P2P energy trading market honestly adhere to optimization protocols and external adversaries are not present, all transmitted real-time data are accurate, enabling the attainment of optimal solutions within finite iterations. Nevertheless, due to the economic-driven or malicious purposes, adversaries may launch cyber-attacks on the market by injecting false data into the transmitted data. Take the malicious purpose for instance, recent years have witnessed notorious events including the 2015 cyber-attacks on Ukrainian power systems \cite{alert2016cyber} and the 2019 cyber-attacks on Venezuela's industrial control systems \cite{bajramovic2019security}. These incidents resulted in national power system blackout and traffic paralysis. Concurrently, there are growing interests in studying cyber-attacks within the context of power systems. Researchers have delved into the impacts of cyber-attacks on volt-var control of power systems \cite{teixeira2014security}, cyber-attacks on real-time electricity markets \cite{choi2013ramp}, small-signal angle stability-oriented false data injections on power systems \cite{hou2022small}, and load redistribution attacks on integrated energy systems \cite{liu2024modeling}. Such malicious attacks can potentially jeopardize the system operation, leading to unreliable energy supplies. To address these concerns, numerous strategies have been proposed to fortify power systems against cyber-attacks, including the prediction interval method \cite{kavousi2020machine}, the population extremal optimization-based deep belief network method \cite{lu2021evolutionary}, the dual deep neural network method \cite{suprabhath2023deep}, and the generalized graph neural network method \cite{takiddin2023generalized}. While these proposed methods have demonstrated to be effective in real-time simulations across various attack scenarios, they are predominantly tailored for centralized power/energy systems.

Comparably, the distributed P2P energy trading market are more vulnerable to cyber-attacks, including the double-spending attack, Sybil attack, and Byzantine attack \cite{tesfamicael2022review}. The double-spending and Sybil attacks primarily target the payment and communication processes in the Blockchain-based P2P energy trading systems \cite{zhang2019double}. In comparison, the Byzantine attack poses a broader and more prevalent cyber-threats in multi-agent P2P energy trading scenarios, often arising from false data injection, replay attacks, or even communication noises \cite{guo2022architecture}. For the Byzantine attack, malicious (Byzantine) agents within the distributed system have the capability of sharing sophisticated false data to neighbors, thereby causing abnormal system states \cite{duan2016resilient,zhao2019resilient}. Meanwhile, extraneous attackers can compromise individual controllers and convert ordinary agents into Byzantine agents \cite{chang2022byzantine}. These vulnerabilities underscore the need for anomaly detection and mitigation mechanisms that can seamlessly integrate with the distributed P2P energy trading scheme to thwart Byzantine attacks.

Recently, researchers have been investigating the cyber-attack mitigation methods for distributed power systems, including the Blockchain consensus method\cite{chen2022blockchain,chen2021distributed}, data-driven method\cite{mohammadi2021detecting,mohiuddin2021deep}, neighbor-observation-based method\cite{duan2016resilient,cheng2021resilient}, and moving average method \cite{chang2022byzantine}. Specifically, the proof of solution mechanism \cite{chen2022blockchain}, and practical Byzantine fault tolerance mechanism \cite{chen2021distributed} are proposed for the distributed energy dispatch and trading. However, these Blochain-based methods necessitate the presence of a reliable coordination committee, which entails multiple rounds of data transmission and verification, thereby compromising the communication efficiency and data privacy. Moreover, the instance-based machine learning classifier \cite{mohammadi2021detecting} and deep learning techniques \cite{mohiuddin2021deep} are proposed to detect false data in the distributed power systems. However, these data-driven methods are not applicable for the realistic P2P energy trading scheme, as the matching of trading partners is random and the convergence trajectory varies at different clearing rounds. Subsequently, some researchers propose the neighbor-observation-based detection method \cite{duan2016resilient,cheng2021resilient}, and moving average method \cite{chang2022byzantine} for the distributed energy management. Nevertheless, the neighbor-observation-based detection method becomes inefficient when encountering sophisticated false data. Similarly, the moving average method becomes inaccurate when the injected false data are large values over multiple iterations. To address the potential cyber-attack towards distributed P2P energy trading market, we propose a physics-informed online spatial-temporal anomaly detection approach. Different from \cite{chen2022blockchain,chen2021distributed}, the proposed approach does not damage the original sparse communication topology and does not impose  additional communication burdens. Meanwhile, compared to \cite{mohammadi2021detecting,mohiuddin2021deep}, the proposed online approach is not limited to varying scenarios. Furthermore, the proposed approach is capable of mitigating sophisticated false data injections, which cannot be addressed by \cite{duan2016resilient,cheng2021resilient,chang2022byzantine}. Overall, we summarize the main contributions as follows:

\begin{itemize}
\item{In contrast to \cite{duan2016resilient,chen2021distributed,chang2022byzantine,liu2023privacy}, this paper develops a fully distributed P2P energy trading model that accounts for the high-fidelity physical network constraints and power losses. Based on the developed model, we investigate adversarial impacts of sophisticated Byzantine faults on the distributed P2P energy trading problem.}
\item{We further propose an online spatial-temporal anomaly detection approach by leveraging the tensor learning method to detect Byzantine faults and mitigate their impacts on the distributed P2P energy trading problem. Specifically, this learning method is informed by the domain knowledge from the developed trading model to enable awesome detection performance. Additionally, we develop closed-form solutions for the proposed detection approach to enhance its computational efficiency.}
\item{We theoretically obtain the optimality and convergence conditions for the distributed P2P energy trading problem (with anomaly detection mechanisms) to attenuate adverse effects from the potentially inaccurate predictions derived by the physics-informed tensor learning method.}
\end{itemize}

The remainder of this paper is organized as follows: Section II introduces the centralized P2P energy trading problem and its distributed reformulations. Section III proposes the physics-informed online anomaly detection approach. Section IV conducts numerical simulations. Section V draws the conclusions.

\section{Formulations of the Real-Time P2P Energy Trading Problem}

\subsection{Centralized P2P Energy Trading Problem}
This paper considers the real-time P2P energy trading problem in radial distribution systems, and its formulation is:

\begin{subequations}
	\begin{align}
		& \mathcal{P}_1: \min\limits_{\mathbf{x}} \quad \sum\limits_{i \in {\cal N}_p} \{ {\alpha _i}\sum\limits_{j \in {\cal S}_i} {e_{i,j}^2} + {\beta _i}\sum\limits_{j \in {{\cal S}_i}} e_{i,j} + {\varepsilon _i} (p_i - p_{i,d})^2 \notag\\
		&\quad\quad\quad + {\omega _b}[ -p_i + \sum\limits_{j \in {\cal S}_i} e_{i,j}]^+ - {\omega _s}[p_i - \sum\limits_{j \in {\cal S}_i} e_{i,j}]^+ \} \label{ob} \\
		&\mathbf{s.t.} \quad  p_i^{\min } \le {p_i} \le p_i^{\max}, i \in {\cal N}_p  \label{box1}\\
		&q_i^{\min } \le {q_i} \le q_i^{\max}, i \in {\cal N}_p \\
		&v_i^{\min} \le {v_i} \le v_i^{\max}, i \in {\cal N}_p \\
		&P_i^{\min} \le {P_i} \le P_i^{\max}, i \in {\cal N}_p \\
		&Q_i^{\min} \le {Q_i} \le Q_i^{\max}, i \in {\cal N}_p \\
		&e_{i,j} \le 0, e_{j,i} \ge 0, i \in {\cal N}_b, j \in {\cal S}_i\\
		&e_{i,j} \ge 0, e_{j,i} \le 0, i \in {\cal N}_s, j \in {\cal S}_i \label{box2}\\
		&e_{i,j} + e_{j,i} = 0, i \in {\cal N}_p, j \in {\cal S}_i  \label{p2p}\\
		&v_{t} - v_i + 2({R_i}{P_i} + {X_i}{Q_i}) - Z_i^2{{\ell}_i}= 0, i \in {\cal N}_p, t \in {\mathcal A}_i \label{gg1} \\
		&\sum\limits_{j \in {\cal C}_i} (P_j-{R_j}{{\ell}_j}) - P_i + p_i = 0, i \in {\cal N}_p \label{gg2}\\
		&\sum\limits_{j \in {\cal C}_i} (Q_j-{X_j}{{\ell}_j}) - Q_i + q_i = 0, i \in {\cal N}_p \label{gg3}\\
		&||2P_i,2Q_i,v_i-\ell_i||_{2} \le v_i+\ell_i, i \in {\cal N}_p \label{gg4}\\
		&p_i - \sum\limits_{j \in {\cal S}_i} e_{i,j} \le 0, i \in {\cal N}_b \label{ban1}\\
		&\sum\limits_{j \in {\cal S}_i} e_{i,j} - p_i \le 0, i \in {\cal N}_s, \label{ban2}
	\end{align}
\end{subequations}
where $[\centerdot]^{+}$ is equivalent to $max\{\centerdot,0\}$. $Z_i^2=R_i^2+X_i^2$, where $R_i$ and $X_i$ are the resistance and reactance of the line connected by the end bus $i$, respectively. In this paper, each bus (node) is considered as an individual prosumer (agent), with the utility company locating at the bus 0. ${\cal N}_p$ is the set for all the nodes in the system, and ${\mathcal A}_i$/${\mathcal C}_i$ are the sets of parent/child nodes for agent $i$. ${\mathcal S}_i$ represents the set of trading partners for prosumer $i$. The objective function \eqref{ob} is to minimize the total operation cost or, equivalently, maximize the social welfare, where clarifications for each term in \eqref{ob} are given in \cite{liu2023privacy}. Vector $\mathbf{x}$ represents the collection of decision variables of all the agents. Specifically, for agent $i$, $\mathbf{x}_i=\{{v _i}, {p _i}, {q _i}, {P _i}, {Q _i}, {{\ell}_i}, \{e_{i,j}\}_{j \in {\cal S}_i}\}$. $v_i/p_i/q_i$ are the voltage magnitude/active power injection/reactive power injection of bus $i$, and $P_i/Q_i/\ell_i$ are the active power flow/reactive power flow/current of the line connected by the end bus $i$. Constraints \eqref{box1}-\eqref{box2} refer to the local boundary constraints for the decision variables. \eqref{p2p} denotes the equality constraint of amounts of traded energy between agents $i$ and $j$, i.e., $e _{i,j}$ and $e _{j,i}$, where $j \in {\mathcal S}_i$. Constraints \eqref{gg1}-\eqref{gg4} represent the second-order cone relaxation of the high-fidelity Disflow model considering power losses for the distribution-level P2P energy trading \cite{zhou2019hierarchical,ullah2021peer}. Power losses have ineligible influences on the P2P decision-making process and results \cite{azim2020investigating}. For the distribution network normally with a tree topology, SOCP relaxation is exact under mild conditions \cite{gan2014exact}. The SOCP relaxation for the relation $\eqref{gg4}$ allows for further derivation of closed-form solutions for the decomposed sub-problems \eqref{sb1} to increase the scalability \cite{peng2016distributed}. Constraints \eqref{ban1} and \eqref{ban2} model the energy balance relations. Agent $i$ is assumed to only be a buyer or a seller at one time slot, depending on its desired active power $p_{i,d}$. Specifically, if $p _{i,d} > 0$, agent $i$ is a buyer and belongs to set ${\mathcal N}_b$. Accordingly, for this case, the constraint \eqref{ban2} is inactive and vice versa. For the case $p _{i,d} = 0$, agent $i$ is not involved in the P2P energy trading, but is still responsible for the power flow calculation.

{\bf Assumption 1:} The cost function in \eqref{ob} for each agent $i$ is denoted as $f_i(\mathbf{x}_i): \mathbb{R}^D \to \mathbb{R}$. $f_i(\mathbf{x}_i)$ is assumed to be  $\delta_i$-strongly-convex as below:
	\begin{align}
		(\mathbf{x}_i-\mathbf{x}_i^*)^T(\nabla f_i(\mathbf{x}_i)-\nabla f_i(\mathbf{x}_i^*)) &\ge \delta_i||\mathbf{x}_i-\mathbf{x}_i^*||_2^2. 
	\end{align}

{\bf Remark 1:} Since each agent $i$ can be either a seller or a buyer at one-time slot, either the forth or fifth term in \eqref{ob} for agent $i$ takes effects. For the case that agent $i$ is no longer involved in the P2P energy trading, its cost function is a constant. Overall, the cost function $f_i(\mathbf{x}_i)$ is convex, and this assumption can be satisfied.

\subsection{Distributed Reformulations of the Trading Problem}
The compact formulation of the centralized P2P energy trading problem $\mathcal{P}_1$ is presented as below:
\begin{subequations}
	\begin{align}
		\mathcal{P}_2: \quad \min\limits_{\mathbf{x}} \quad & \sum\limits_{i \in {\cal N}_p} f_i(\mathbf{x}_i) \label{cp1}\\
		\mathbf{s.t.} \quad & \sum\limits_{j \in {\cal C}_i\cup{\cal A}_i\cup{\cal S}_i}H_{i,j}\mathbf{x}_i=0, \quad i \in {\cal N}_p \label{cp2}\\
		&\mathbf{x}_i \in \mathscr{X}_i, \quad i \in {\cal N}_p, \label{cp3}
	\end{align}
\end{subequations}
where $H_{i,j}$ is the coefficient matrix for the collection of global constraints \eqref{p2p}-\eqref{gg3}. $\mathscr{X}_i$ is the convex set for each agent $i$, which is related to the local constraints \eqref{box1}-\eqref{box2}, and \eqref{gg4}-\eqref{ban2}.
The compact form of P2P energy trading problem, $\mathcal{P}_2$, can be solved in a fully distributed manner by only sharing the values of partial decision variables with neighboring nodes and P2P trading partners. To this end, a set of auxiliary variables $\mathbf{y}_{i(j)}$ is introduced for duplicating the variable $\mathbf{x}_{i}$ for each agent $i$. Accordingly, $\mathcal{P}_2$ is reformulated to $\mathcal{P}_3$ as below:
\begin{subequations}
\begin{align}
		\mathcal{P}_3: \quad  \min\limits_{\mathbf{x,y}}& \quad  \sum\limits_{i \in {\cal N}_p} f_i(\mathbf{x}_i) \label{dp1}\\
		\mathbf{s.t.} \quad & \sum\limits_{j \in {\cal C}_i\cup{\cal A}_i\cup{\cal S}_i} H_{i,j}\mathbf{y}_{i(j)}=0, \quad i \in {\cal N}_p \label{dp2}\\
		&\mathbf{x}_i \in \mathcal{X}_i, \quad i \in {\cal N}_p \label{dp3} \\
    	&\mathbf{x}_i - \mathbf{y}_{i(j)} =0, \quad i \in {\cal N}_p, j \in {\cal C}_i/{\cal A}_i/{\cal S}_i,  \label{dp4}
\end{align}
\end{subequations}
where $\eqref{dp4}$ denotes the equality constraints for $\mathbf{x}_{i}$ and $\mathbf{y}_{i(j)}$.
Let $\mathbf{\mu}_{i(j)}$ be the vector of dual variables with regard to the constraint \eqref{dp4}. The decomposed augmented Lagrangian function for each agent $i$ can be derived as below:
\begin{subequations}
\begin{align} 
 		&\mathcal{L}_i(\mathbf{x}_i, \mathbf{y}_{i(j)},\mathbf{\mu}_{i(j)})  \notag\\
 		&=\resizebox{0.84\hsize}{!}{$f_i(\mathbf{x}_i)+\sum\limits_{j \in {\cal C}_i\cup{\cal A}_i\cup{\cal S}_i}<\mathbf{\mu}_{i(j)}, \mathbf{x}_i-\mathbf{y}_{i(j)}>+\frac{\eta_i}{2}||\mathbf{x}_i-\mathbf{y}_{i(j)}||_2^2 $} \\
 		&=\resizebox{0.84\hsize}{!}{$ f_i(\mathbf{x}_i)+\sum\limits_{i \in {\cal C}_j\cup{\cal A}_j\cup{\cal S}_j}<\mathbf{\mu}_{j(i)}, \mathbf{x}_j-\mathbf{y}_{j(i)}>+\frac{\eta_j}{2}||\mathbf{x}_j-\mathbf{y}_{j(i)}||_2^2,$}
\end{align}
\end{subequations}
where $<,>$ represents the inner product. $\eta_i$ and $\eta_j$ are the penalty factors for agents $i$ and $j$, respectively.
The augmented Lagrangian problem can be further solved iteratively via the alternating direction method of multipliers (ADMM), which consists the following steps: 
 \begin{subequations}
	\begin{align}
		&\mathbf{x}_i^{k+1}= \mathop{\text{arg min}}\limits_{\mathbf{x}_i \in \mathcal{X}_i} \mathcal{L}_i(\mathbf{x}_i, \mathbf{y}_{i(j)}^{k},\mathbf{\mu}_{i(j)}^{k}) \label{sb1}\\
		&\mathbf{y}_{i(j)}^{k+1}= \mathop{\text{arg min}}\limits_{\mathbf{y}_i \in \mathcal{Y}_i} \mathcal{L}_i(\mathbf{x}_i^{k+1}, \mathbf{y}_{i(j)},\mathbf{\mu}_{i(j)}^{k}) \label{sb2} \\
		&\mathbf{\mu}_{i(j)}^{k+1}= \mathbf{\mu}_{i(j)}^{k}+\rho_i(\mathbf{x}_i^{k+1}-\mathbf{y}_{i(j)}^{k+1}),  \label{sb3}
	\end{align}
\end{subequations}
where $\mathcal{Y}_i = \{\mathbf{y}_{i(j)}|\sum\limits_{j \in {\cal N}_i} H_{i,j}\mathbf{y}_{i(j)}=0\}$  and ${\cal N}_i={\cal C}_i\cup{\cal A}_i\cup{\cal S}_i$. In this work, closed-form solutions for \eqref{sb2} are compulsory for deriving convergence conditions of the distributed P2P energy trading problem, and thus we provide corresponding details. Closed-form solutions for the sub-problems \eqref{sb1} are also accessible and we refer interested readers to \cite{peng2016distributed,li2018distributed} for details. The compact form for problem \eqref{sb2} is as follows:
 \begin{subequations}
	\begin{align}
		\mathcal{P}_4: \quad  \min\limits_{\mathbf{R}_i} \quad&  \frac{1}{2}\mathbf{R}_i^T\mathbf{G}_i\mathbf{R}_i+v^T\mathbf{R}_i  \label{close1}\\
        \mathbf{s.t.} \quad&  \mathbf{M}_i\mathbf{R}_i=0, \label{close1}
	\end{align}
\end{subequations}
where $\mathbf{M}_i$ is the positive diagonal matrix for the coupling constraints. $\mathbf{G}_i$ is the real-value matrix and $v$ is the coefficient vector, i.e., $v=-\mathbf{\mu}_{i(j)}^{k}-\eta_i\mathbf{x}_i^{k+1}$. $\mathbf{R}_i$ is the decision variables, i.e., $\mathbf{y}_{i(j)}^{k+1}$. Then closed-from solutions for problem $\mathcal{P}_4$ are:
	\begin{align}
		\mathbf{R}_i=(\mathbf{G}_i^{-1}\mathbf{M}_i^{T}(\mathbf{M}_i\mathbf{G}_i^{-1}\mathbf{M}_i^{T})^{-1}\mathbf{M}_i\mathbf{G}_i^{-1}-\mathbf{G}_i^{-1})v,  \label{close3}
	\end{align}
where the solutions can be rewritten as $\mathbf{y}_{i(j)}^{k+1}=\mathcal{Q}_i \cdot(-\mathbf{\mu}_{i(j)}^{k}-\eta_i\mathbf{x}_i^{k+1})$ and $\mathcal{Q}_i$ is the squeezed coefficient matrix before variable $v$ in \eqref{close3}. Conversely, dual variables $\mathbf{\mu}_{i(j)}^{k}$ can be represented as $\mathbf{\mu}_{i(j)}^{k}=-\mathcal{Q}_i^{-1}\mathbf{y}_{i(j)}^{k+1}-\eta_i\mathbf{x}_i^{k+1}$, which is crucial for deriving the convergence conditions in Section III.D.

\section{Spatial-temporal Anomaly Detection via Online Tensor Learning}
\subsection{Distributed Data-Sharing Scheme}
\begin{figure}[!tbp]
	\centering	
	\includegraphics[width=1.0\linewidth]{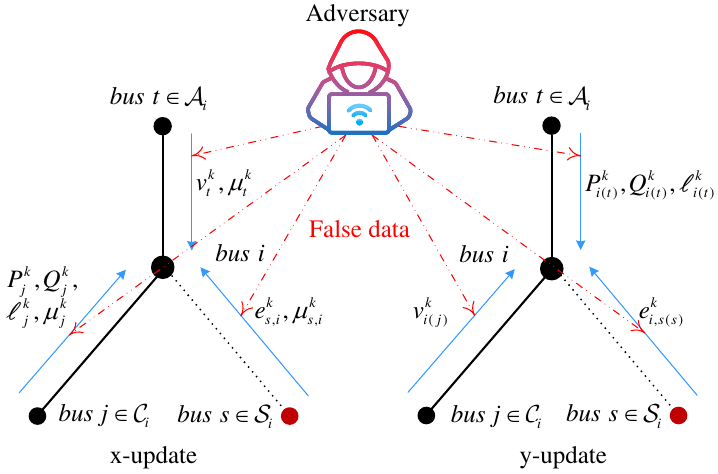}
	\caption{False data injections during distributed optimization.}
    \vspace{-0.5cm}
	\label{fig1}
\end{figure}

In the distributed P2P energy trading problem ${\cal P}_3$, each agent shares partial data with its neighbors and trading partners iteratively. As illustrated in Fig. \ref{fig1}, during the x-update at iteration $k$, agent $i$ (at the bus $i$) receives real-time data from other agents, i.e., $\{\{{v_t^k,\mu_t^k}\}_{t \in {\cal A}_i},$ $ \{{P_j^k}, {Q_j^k},{{\ell}_j^k},\mu_j^k\}_{j \in {\cal C}_i},$ $ \{e_{s,i}^k,\mu_{s,i}^k\}_{s \in {\cal S}_i}\}$. Meanwhile, during the y-update at iteration $k$, agent $i$ receives data from other agents, i.e., $\{\{{P_{i(t)}^k}, {Q_{i(t)}^k}, {{\ell}_{i(t)}^k}\}_{t \in {\cal A}_i},$ $\{v_{i(j)}^k\}_{j \in {\cal C}_i},$ $ \{e_{i,s(s)}^k\}_{s \in {\cal S}_i}\}$. While the decentralized optimization is well-established for faultless networks, this paper assumes that some network nodes may arbitrarily deviate from their intended behavior during the iterative process. Specifically, received data for each agent can deviate from their true values, as they can be compromised by malicious agents or extraneously adversaries through injecting sophisticated false data or equipment malfunctions, termed as Byzantine faults. These false data can cause infeasibility or divergence of the distributed optimization methods, such as the ADMM. To address this, we here model the node with deviations, which is formally defined as follows.
	
{\bf Definition 1:} (Byzantine agent) \cite{fang2022bridge}. Agent $i, i \in \mathcal{N}_p, $ is defined to have undergone a Byzantine failure if, during any iteration of the distributed optimization, it either updates its local variable $\mathbf{x}_i$ without following the update rule $\eqref{sb1}-\eqref{sb3}$ or it broadcasts some information other than its true local information to other nodes in its vicinity.

The Byzantine-resilient concept, used in distributed systems and computer science \cite{lamport2019byzantine}, refers to a system's ability to function correctly and reliably despite arbitrary faults or malicious behaviors. A system is considered Byzantine-resilient if it can preserve its integrity, correctness, and functionality even in the presence of a certain number of Byzantine nodes within the network. The maximum allowable number of such nodes is typically determined by the specific algorithm or protocol employed. To mitigate these impacts and ensure the Byzantine-resilience of the real-time P2P energy trading problem, we propose an online spatial-temporal anomaly detection approach, which is informed by the domain knowledge from the distributed P2P energy trading model. We further develop closed-form solutions for updating model parameters to enhance its computational efficiency, and subsequently provide theoretical optimality and convergence conditions for the distributed P2P energy trading problem (with the proposed anomaly detection approach) to attenuate adverse effects from the potential occurrence of unreliable predictions.
\vspace{-0.1cm}

\subsection{Online Spatial-Temporal Anomaly Detection Mechanism}
\begin{figure*}[!tbp]
	\centering
	\includegraphics[width=1.0\linewidth]{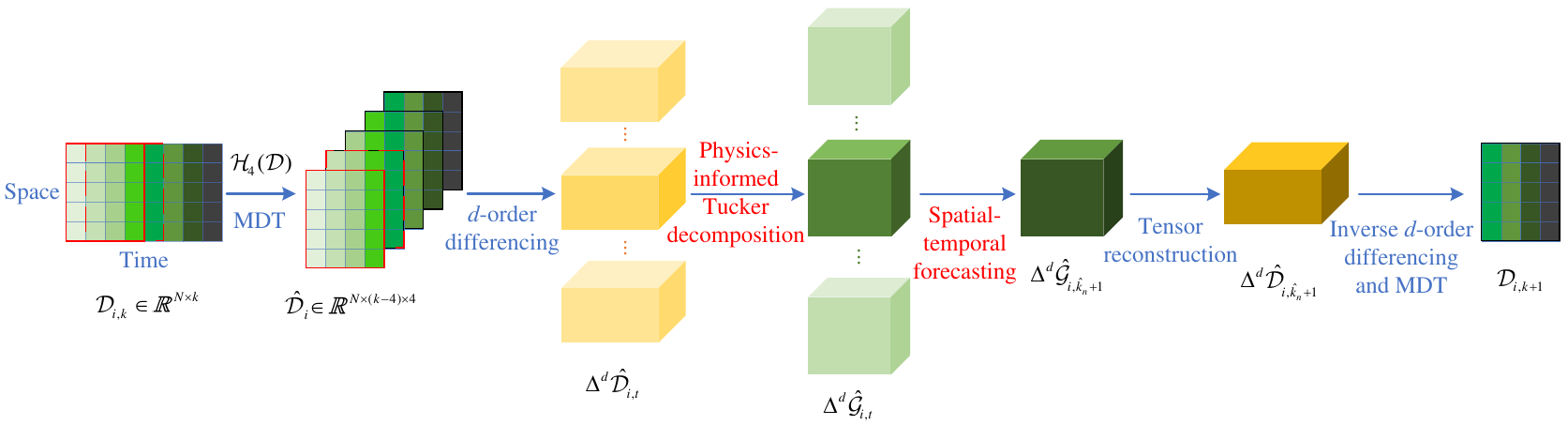}
	\caption{Online spatial-temporal anomaly detection via physics-informed tensor learning.}
	\label{fig2}
\end{figure*}

In fact, iteratively shared data for each agent in the distributed optimization are spatial-temporally correlated. Considering that data-sharing schemes are homogeneous for the x-update and y-update, we take the y-update as an example. At each iteration, received data from other agents, allied with personal data, are stored locally by each agent to form a streaming dataset. This dataset has temporal properties. Normally, each kind of data in this dataset, e.g., $v_i^k$, is an univariate time series containing temporal autocorrelations \cite{santer2000statistical}, as $v_i^k$ is gradually updated over iterations via the ADMM. Meanwhile, at each iteration, this dataset is also spatially correlated by the global constraints, i.e., \eqref{p2p}-\eqref{gg4}. By leveraging these properties, we propose a physics-informed online learning method as in Algorithm 1 to detect anomaly of the streaming dataset for each agent. This algorithm further contributes to the Byzantine-resilient distributed algorithm as denoted in Algorithm 2. To better elucidate the concepts embedded within Algorithm 1, a visual depiction of its logic is given in Fig. \ref{fig2}. 

\begin{algorithm}[!htbp]
	\caption{Online Physics-Informed Tensor Learning.}\label{alg1}
	\begin{algorithmic}
		\STATE 
		\STATE {\textbf{Initialization}}
		\STATE \hspace{0.5cm}$ \text{Initialize } \tau, p, d,q,\hat{k}_n, h, H, n, \tau_n, N,\hat{\mathbf{A}}_i^{(n)},\mathbb{\hat{\epsilon}}_{t-i}^{(n)}, \xi$ 
		\STATE \hspace{0.5cm}$ \text{Obtain } \mathcal{\hat{D}}_{i}, \{\Delta^d \mathcal{\hat{D}}_{i,t}\}_{t=d}^{\hat{k}_n} \text{ from } \mathcal{D}_{i,k}$ following \eqref{close4}
		\STATE {\textbf{Spatial-Temporal Tucker Decomposition}}
		\STATE \hspace{0.5cm}$ \textbf{for } h \le H \textbf{ do}$ 
		\STATE \hspace{0.8cm}$ \text{Compute } \{ \resizebox{0.73\hsize}{!}{$\Delta^d\mathcal{\hat{G}}_{i,t}= \Delta^d\mathcal{\hat{D}}_{i,t} \times_1 \hat{\mathbf{A}}_i^{(1)^T}\cdots  \times_N \hat{\mathbf{A}}_i^{(N)^T}\}_{t=1}^{\hat{k}_n}$} $
		\STATE \hspace{0.8cm}$ \text{Compute coefficients } \psi_m, \theta_m \text{ by Yule-Walker equations}$
		\STATE \hspace{0.8cm}$ \textbf{for } n \le N \textbf{ do}$ 
		\STATE \hspace{1.5cm} $\text{Update } \Delta^d\mathcal{\hat{G}}_{i,t}^{(n)}  \text{ following } \eqref{up1}$ 
		\STATE \hspace{1.5cm} $\text{Update } \{\mathbb{\hat{\epsilon}}_{i,t-m}^{(n)}\}_{m=1}^q   \text{ following } \eqref{up2}$
		\STATE \hspace{1.5cm} $\text{Update } \hat{\mathbf{A}}_i^{(n)^h} \text{ following } \eqref{up3}$
		\STATE \hspace{1.5cm} $\text{Obtain } \Delta^d\mathcal{\hat{G}}_{i,t}, \mathbb{\hat{\epsilon}}_{i,t-m} \text{ by folding }  \Delta^d\mathcal{\hat{G}}_{i,t}^{(n)}, \mathbb{\hat{\epsilon}}_{i,t-m}^{(n)}$	 
		\STATE \hspace{0.8cm}\textbf{end}
		\STATE \hspace{0.8cm}\textbf{if } $\sum_{n=1}^N ||\hat{\mathbf{A}}_i^{(n)^{h+1}}-\hat{\mathbf{A}}_i^{(n)^h}||_F^2 \le \xi \cdot \sum_{n=1}^N ||\hat{\mathbf{A}}_i^{(n)^h}||_F^2$ 
   		\STATE \hspace{1.5cm} \textbf{break}
		\STATE \hspace{0.8cm}\textbf{else}
		\STATE \hspace{1.5cm} $h \leftarrow h+1$ 	
		\STATE \hspace{0.8cm}\textbf{end} 	
		\STATE \hspace{0.5cm}\textbf{end}   
		\STATE {\textbf{Physics-Informed Online Forecasting}}
		\STATE \hspace{0.8cm}$ \text{Compute } \Delta^d \mathbf{\mathcal{\hat{G}}}_{i,\hat{k}_n+1} \text{ following } \eqref{new1} $
		\STATE \hspace{0.8cm}$ \text{Compute } \Delta^d\mathcal{\hat{D}}_{i,\hat{k}_n+1} \text{ following } \eqref{new2} $
		\STATE \hspace{0.8cm}$ \text{Obtain } \mathbf{\mathcal{D}}_{i,k+1} \text{ following } \eqref{new3} $
		\STATE \hspace{0.8cm}$ \text{Calculate } \gamma_i^{k+1} \text{ following } \eqref{ref5},\eqref{ref6} $
		\STATE \textbf{Return} $\gamma_i^{k+1}$	
	\end{algorithmic}
\end{algorithm}

As illustrated in Fig. \ref{fig2}, for each agent $i$ at the $(k+1)$-th iteration, the streaming dataset $\mathcal{D}_{i,k}$, $\mathcal{D}_{i,k} \in \mathbb{R}^{I_1\cdots \times I_N}$ and $N=2$, can be first constructed by employing historical received and personal data (from the (k-L)-th to k-th iterations). $\mathcal{D}_{i,k}$ is further transformed into high-order tensors through multi-way delay embedding transform (MDT) over $N$ modes:   
\begin{align}
	\mathcal{\hat{D}}_i=\mathcal{H}_{\tau}(\mathcal{D}_{i,k})=Fold_{I,\tau}(\mathcal{D}_{i,k}\times_{1}\mathbf{S}_{1}\cdots\times_{N}\mathbf{S}_{N}), \label{close4}
\end{align}
where $\mathcal{\hat{D}}_i \in \mathbb{R}^{J_1\cdots \times J_N}$ is high-order block Hankel tensor, which is assumed to be low-rank. Let $J_n= \tau_n\times \hat{k}_n$ and $ \hat{k}_n=(k-\tau_n+1)$. $\mathbf{S}_{n} \in \mathbb{R}^{J_n\times I_n}$ is the duplication matrix, and $Fold_{I,\tau}$ is the mapping operator: $\mathbb{R}^{J_1 \cdots \times J_N} \longrightarrow \mathbb{R}^{\prod_n(\tau_n) \times \prod_n(I_n-\tau_n+1)}$. To avoid heavy computational burden, MDT is only employed along the temporal mode to capture temporal correlations. Let the $N$-th dimension represents the temporal mode:
\begin{align}
	\mathcal{\hat{D}}_i=\mathcal{H}_{\tau}(\mathcal{D}_{i,k}) \in \mathbb{R}^{J_1\cdots \times J_N},
\end{align}
where $J_n=I_n$ for $n=1,...,N-1$, and $J_N=\tau_n\times \hat{k}_n$. 
To leverage the high-order data to predict data at the next iteration,  we develop an online physics-informed anomaly detection mechanism by incorporating the spatial-temporal Tucker decomposition method into the auto-regressive integrated moving average (ARIMA).

\begin{figure}[!tbp]
	\centering
	\includegraphics[width=1.0\linewidth]{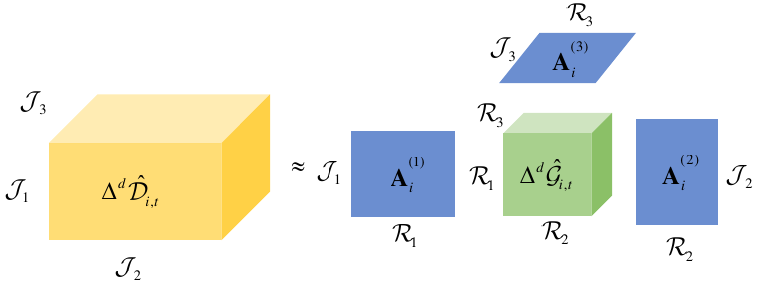}
	\caption{Scheme of physics-informed Tucker decomposition.}
	\label{fig3}
\end{figure}

Let $\{\Delta^d \mathcal{\hat{D}}_{i,t}\}_{t=d}^{\hat{k}_n}$ be the $d$-order differencing of the block tensor $\mathcal{\hat{D}}_i$. As denoted in Fig. \ref{fig3}, the differenced tensor can be decomposed via the Tucker decomposition method \cite{shi2020block}:  
\begin{subequations}
	\begin{align}
		\mathcal{P}_5: \quad \Delta^d\mathcal{\hat{D}}_{i,t} &= \Delta^d\mathcal{\hat{G}}_{i,t}\times_1 \hat{\mathbf{A}}_i^{(1)} \cdots  \times_N \hat{\mathbf{A}}_i^{(N)}  \\
		\mathbf{s.t.} &\quad  \hat{\mathbf{A}}_i^{(n)^T}\hat{\mathbf{A}}_i^{(n)}=\mathbf{I}, n=1,...,N,
	\end{align}
\end{subequations}
where $\Delta^d\mathcal{\hat{G}}_{i,t} \in \mathbb{R}^{R_1\times R_2 \times \cdots R_N}$ is the low-rank core tensor. The Tucker-rank of $\Delta^d\mathcal{\hat{G}}_{i,t}$ is the $N$-dimensional vector: $[R_1,...,R_n,...,R_N]$, and $R_n$ is the rank of the mode-$n$ unfolded matrix $\Delta^d\mathcal{D}_{i,t}$. The projection matrix $\hat{\mathbf{A}}_i^{(n)}, \hat{\mathbf{A}}_i^{(n)} \in \mathbb{R}^{J_n \times R_n}$, is orthogonal and non-negative. The operator $\times_n$ represents the mode-$n$ matrix product of a tensor \cite{kolda2009tensor}.

The $(p,d,q)$-order ARIMA model is employed to capture correlations of core tensors as:
\begin{align}
	\Delta^d\mathcal{\hat{G}}_{i,t}=\sum_{m=1}^p \psi_m \Delta^d\mathcal{\hat{G}}_{i,t-m} -\sum_{m=1}^q \theta_m \mathbb{\hat{\epsilon}}_{i,t-m}+\mathbb{\hat{\epsilon}}_{i,t},
\end{align}
where $\psi_m$ and $\theta_m$ are coefficients inside ARIMA, which represent orders of the lag and moving average, respectively. $\mathbb{\hat{\epsilon}}_{i,t}$ denotes the forecast error at $t$-time step for agent $i$.

The physical spatial correlation between multiple series data, $\Delta^d\mathcal{\hat{D}}_{i,t},$ is modeled as below:
\begin{align}
	\Delta^d\mathcal{\hat{D}}_{i,t} \mathcal{W}_i =\Delta^d\mathcal{\hat{G}}_{i,t} \prod_{n=1}^{N} \times_{n} \hat{\mathbf{A}}_i^{n}\mathcal{W}_i = 0, \label{sp}
\end{align}
where $\mathbf{\mathcal{W}}_i$ is the local spatial correlation matrix for agent $i$, i.e., the node admittance matrix. The original constructed multiple series data for each agent $i$, i.e., $\mathcal{D}_{i,k}$, are spatially correlated, which are directly related to the constraints \eqref{p2p}-\eqref{gg3} in $\mathcal{P}_1$. Ideally, these relations indicate $\mathcal{D}_{i,k}\mathcal{W}_i=0$. Through the linear transformations, i.e., MDT and $d-$order differencing, $\Delta^d\mathcal{\hat{D}}_{i,t},$ are still spatially correlated, which indicates, $\Delta^d\mathcal{\hat{D}}_{i,t} \mathcal{W}_i=0$. This correlation is imposed through the domain knowledge, which ensure that the learned core tensor, i.e., $\Delta^d\mathcal{\hat{G}}_{i,t}$, also admit the equivalent correlations, i.e., $\Delta^d\mathcal{\hat{G}}_{i,t} \prod_{n=1}^{N} \times_{n} \hat{\mathbf{A}}_i^{n}\mathcal{W}_i = 0$. To capture these properties, equation \eqref{sp} is thus established. These relations are further integrated with the Tucker decomposition and ARIMA learning process, forming the physics-informed online tensor learning approach as below.
\begin{subequations}
	\begin{align}
		\mathcal{P}_6: & \min\limits_{\substack{\Delta^d\mathcal{\hat{G}}_{i,t}^{(n)}, \\ \hat{\mathbf{A}}_i^{(n)}, \mathbb{\hat{\epsilon}}_{i,t-m}^{(n)}, \\ \psi_m, \theta_m}} \sum\limits_{\substack{t=p+d\\+q+1}}^{\hat{k}_n}\sum_{n=1}^{N}(\frac{1}{2} ||\Delta^d\mathcal{\hat{G}}_{i,t}^{(n)}-\hat{\mathbf{A}}_i^{(n)^T}\Delta^d\mathcal{\hat{D}}_{i,t}^{(n)}\hat{\mathbf{A}}_i^{(-n)^T}||_F^2  \notag \\ 
		&+\frac{1}{2}||\Delta^d\mathcal{\hat{G}}_{i,t}^{(n)}-\sum_{m=1}^p \psi_m \Delta^d\mathcal{\hat{G}}_{i,t-m}^{(n)} + \sum_{m=1}^q \theta_m \mathbb{\hat{\epsilon}}_{i,t-m}^{(n)}||_F^2 \notag \\
		& +\frac{1}{2} ||\hat{\mathbf{A}}_i^{(n)}\Delta^d\mathcal{\hat{G}}_{i,t}^{(n)}\hat{\mathbf{A}}_i^{(-n)} \mathcal{W}_i||_F^2) \label{op1} \\
    	\mathbf{s.t.} &\quad  \hat{\mathbf{A}}_i^{(n)^T}\hat{\mathbf{A}}_i^{(n)}=\mathbf{I}, n=1,...,N, \label{op2} 
	\end{align}
\end{subequations}
where the objective function contains three parts: i) the minimization of the Tucker decomposition error; ii) the minimization of the forecast error of ARIMA  model; iii) the minimization of the spatial correlated error, which is imposed through the domain knowledge. Different weight parameters for these parts can have influences on the overall performance of the proposed approach. To balance influences of different parts, we employ a classic equal weight, i.e., $\frac{1}{2}$, for each part in the objective function in \eqref{op1} via the least square formulation. $\hat{\mathbf{A}}_i^{(-n)}=\hat{\mathbf{A}}_i^{(N)^T} \otimes \cdots \hat{\mathbf{A}}_i^{(n+1)^T}\otimes \hat{\mathbf{A}}_i^{(n-1)^T} \cdots \hat{\mathbf{A}}_i^{(1)^T} \in \mathbb{R}^{\prod_{j \neq n} R_j \times \prod_{j \neq n} I_j}$, which is a constant. The third term in the objective function can be simplified as $\frac{1}{2} ||\Delta^d\mathcal{\hat{G}}_{i,t}^{(n)}\hat{\mathbf{A}}_i^{(-n)} \mathcal{W}_i||_F^2$ due to the orthogonal property of $\hat{\mathbf{A}}_i^{(n)} $. Each agent $i$ can conduct this physics-informed decomposition for anomaly detection, and closed-form solutions are accessible for updating the decision variables. Details of closed-form solutions can be found in Appendix A.

\subsection{Anomaly Detection via Online Tensor Learning}
The new predicted core tensor and reconstructed dataset at the $(k+1)$-th iteration can thus be obtained:
\begin{subequations}
\begin{align}
	&\Delta^d \mathbf{\mathcal{\hat{G}}}_{i,\hat{k}_n+1} = \sum_{m=1}^p \psi_m \Delta^d\mathcal{\hat{G}}_{i,\hat{k}_n-m}- \sum_{m=1}^q \theta_m \mathbb{\hat{\epsilon}}_{i,\hat{k}_n-m} \label{new1}\\
	
	&\Delta^d\mathcal{\hat{D}}_{i,\hat{k}_n+1} = \Delta^d\mathcal{\hat{G}}_{i,\hat{k}_n+1} \prod_{n=1}^N \times_n \hat{\mathbf{A}}_i^{(n)} \label{new2}\\
	
	&\mathbf{\mathcal{D}}_{i,k+1} = \mathcal{H}_{\tau}^{-1}({\mathcal{\hat{D}}_{i,\hat{k}+1}}) \in \mathbb{R}^{I_1\cdots I_{N-1}\times I_N\times (k+1)}. \label{new3}
 
\end{align}
\end{subequations}
The data at the $(k+1)$-th iteration predicted by the online tensor learning is denoted as $d_{i,k+1}^p$, which is the $(k+1)$-th slice from the dataset $\mathbf{\mathcal{D}}_{i,k+1}$. Received data at the $(k+1)$-th iteration is denoted as $d_{k+1}^r$. The anomaly of shared data for agent $i$ is detected by the following criteria:
\begin{subequations}
	\begin{numcases}
		{\gamma_i^{k+1}=}     
		0,  \enspace if\enspace ||d_{i,k+1}^r-d_{i,k+1}^p||_2 \le \varphi_i \label{ref5} \\
		1, \enspace otherwise, \label{ref6}  
	\end{numcases}
\end{subequations}
where $\gamma_i^{i,k+1}$ is the false alarm rate. When the anomaly is detected, i.e., $\gamma_i^{i,k+1}=1$, the update for $d_{i,k+1}$ follows:
\begin{subequations}
	\begin{numcases}
		{d_{i,k+1}=}     
		d_{i,k+1}^{p},  \enspace if\enspace ||d_{i,k+1}^p-d_{i,k}||_2 \notag\\
		\quad\quad\quad\quad \le \lambda_i||d_{i,k}-d_{i,k-1}||_2  \label{ref7} \\
		d_{i,k}, \enspace otherwise.   \label{ref8}
	\end{numcases}
\end{subequations}
When $d_{i,k+1}=d_{i,k}$ happens, the predicated value $d_{i,k+1}^p$ derived by the proposed online physics-informed tensor learning method is deemed as relatively inaccurate. Otherwise, i.e., $\gamma_i^{k+1}=0$, $d_{i,k+1}$ is updated as:
\begin{align}
	d_{i,k+1} = d_{i,k+1}^r. \label{ref9}
\end{align}

Data $d_{i,k+1}$ can be utilized to update the primal or dual variables for agent $i$, i.e., $\mathbf{x}_i^{k+1}$ and $\mathbf{y}_{i(j)}^{k+1}$. Therefore, we develop a Byzantine-resilient distributed algorithm as denoted in Algorithm 2. Meanwhile, we can define primal and dual residuals at the $(k+1)$-th iteration for the Byzantine-resilient distributed algorithm in Algorithm 2 as:
\begin{align}
		\resizebox{0.88\hsize}{!}{$
		\gamma_p^{k+1}=\sum\limits_{i \in {\cal N}_p}||\mathbf{x}_i^{k+1}-\mathbf{y}_{i(j)}^{k+1}||_2,	 \gamma_d^{k+1}=\sum\limits_{i \in {\cal N}_p}\eta_i||\mathbf{y}_{i(j)}^{k+1}-\mathbf{y}_{i(j)}^{k}||_2$},
\end{align}
where $j \in {\cal C}_i\cup{\cal A}_i\cup{\cal S}_i$ for each agent $i$.

\begin{algorithm}[!htbp]
	\caption{Byzantine-Resilient Distributed Algorithm.}\label{alg2}
	\begin{algorithmic}
		\STATE 
		\STATE {Initialize all the primal, auxiliary, and dual variables}
		\STATE $ \textbf{While } \gamma_p^k \ge \varpi_1 \text{ and } \gamma_p^k \ge \varpi_2  \text{ , } k \le k_{max} \textbf{ do}$
		\STATE \hspace{0.5cm}$ \textbf{For} \text{ each agent } i \in {\cal N}_p \textbf{ do}$
		\STATE \hspace{0.8cm}$ \text{Update primal variables } \mathbf{x}_i^{k} \text{ following \eqref{sb1}} $
		\STATE \hspace{0.8cm}$ \text{Share variables following the routine denoted in Fig.\ref{fig1}} $
		\STATE \hspace{0.8cm}$ \text{Construct dataset } \mathcal{D}_{i,k} \text{ via received and personal data}$
		\STATE \hspace{0.8cm}$ \text{Detect data anomaly following Algorithm \ref{alg1}} \text{ using } \mathcal{D}_{i,k}$
		\STATE \hspace{0.8cm}\textbf{if } $\gamma_i^{k}=1$
		\STATE \hspace{1.1cm}$ \text{Update variables } d_{i,k} \text{ following \eqref{ref7}}, \eqref{ref8} $
		\STATE \hspace{0.8cm}\textbf{else}
		\STATE \hspace{1.1cm}$ \text{Update variables } d_{i,k} \text{ following \eqref{ref9}}$	
		\STATE \hspace{0.8cm}\textbf{end}
		\STATE \hspace{0.8cm}$ \text{Update auxiliary variables } \mathbf{y}_i^{k} \text{ following } \eqref{sb2} $	
		\STATE \hspace{0.8cm}$ \text{Update dual variables } \mathbf{\mu}_{i(j)}^{k} \text{ following \eqref{sb3}} $	
		\STATE \hspace{0.5cm}$ \textbf{end} $
		\STATE \hspace{0.5cm}$ k \leftarrow k+1 $	
		\STATE $\textbf{end}$
	\end{algorithmic}
\end{algorithm}

\subsection{Optimality and Convergence Conditions}
In this subsection, we analyze and derive the optimality and convergence conditions of Algorithm 2 to mitigate effects from potentially inaccurate predictions derived by the proposed online physics-informed tensor learning method. Let $\Upsilon$ be the percentage of prosumers who always employ the true or predicted values over iterations, and $1-\Upsilon$ represents the percentage of prosumers who encounter unreliable predictions.

{\bf Theorem 1}:  The proposed Byzantine-resilient distributed algorithm for the P2P energy trading problem can converge to the stationary solution if the following conditions are satisfied:
\begin{align}
	&\frac{\Upsilon\eta_i}{2}-\frac{2\mathcal{Q}_i^{-2} \lambda_i^2}{\eta_i}>0, \quad \frac{\Upsilon(\delta_i+\eta_i)}{2}-2\eta_i\lambda_i^2 >0. \label{cond1}
\end{align}	
Proof: We provide the detailed proof in the Appendix B.

\section{Numerical Simulations}
\subsection{Simulation Setup}
The code is implemented in Python 3.9 on a computer with 6-core Intel i5-10500 CPU@3.10GH, and the state-of-the-art Gurobi 11.0 is employed to solve the decomposed sub-problems. The standard IEEE 15-bus and 85-bus distribution systems \cite{zimmerman2010matpower} are employed to verify the effectiveness, optimality, and scalability of the Byzantine-resilient distributed algorithm with the proposed anomaly detection approach for the real-time P2P energy trading problem. The real PV and wind power generation data are derived from \cite{jenkins2019hybrid}. Without losing generality, parameters of the utility function and the discomfort cost function, are decided based on the existing literature \cite{ullah2021peer} and our previous research \cite{liu2023online}, which are made random for realistic purposes. Specifically, the following parameters are randomly generated for each prosumer: i) the parameters of the utility function for buyers are $\alpha _i$, $\alpha _i \in [0.01,0.1]$ ¢/kWh$^2$, and $\beta_i$, $\beta_i \in [1.0,3.0]$ ¢/kWh; ii) the corresponding parameters for sellers are $\alpha _i$, $\alpha _i \in [0.02,0.1]$ ¢/kWh$^2$, and $\beta_i$, $\beta_i \in [0.1,0.8]$ ¢/kWh; iii) the parameter for the discomfort cost function is $\varepsilon _i$, $\varepsilon _i \in [2.5,3.5]$ ¢/kWh$^2$. For other hyper-parameters, i.e., $p, d, q, H, R, L,$ and etc., we firstly narrowed down their ranges based on the empirical experiences and existing literature \cite{shi2020block}, and finally decided the appropriate combination of hyper-parameters according to acceptable results from the experimental outcomes. The stopping criteria for primal and dual residuals are set as $\varpi_1 = 10^{-4}$ and $\varpi_2=10^{-4}$. The penalty factor $\eta_i$ is set to 1 for all the agents. The detection parameter $\varphi_i$ is set to 0.1, and the $\lambda_i$ is set to 1.0. The window length for constructing the streaming dataset $\mathcal{D}_{i,k}$ for each agent $i$ is set to be 30, i.e., $L=30$, to balance the computational efficiency and detection performance. For the proposed physics-informed online tensor learning method, the parameters are set as $p=2, d=2, q=1, H=10, R=4, \xi=10^{-4}$. We concede that these suggested hyper-parameters might not be the combination leading to the optimal overall performance, while there could be more systematic approaches for deciding the optimal combination, including the promising grid search technique, nature-inspired technique, and gradient descent method \cite{khalid2020survey, alsobhi2022prediction}.
\begin{figure}[!hbp]
	\centering
	\includegraphics[width=1.02\linewidth]{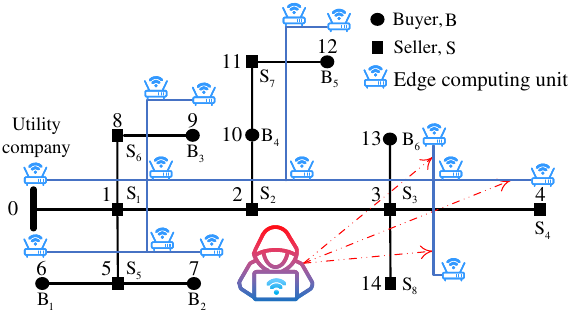}
	\caption{IEEE 15-bus systems with Byzantine faults.}
	\label{fig4}
	\vspace{-0.5cm}
\end{figure}

\subsection{Correlated Static False Data Injections}
\begin{figure}[htbp]
	\centering
	\includegraphics[width=1.0\linewidth]{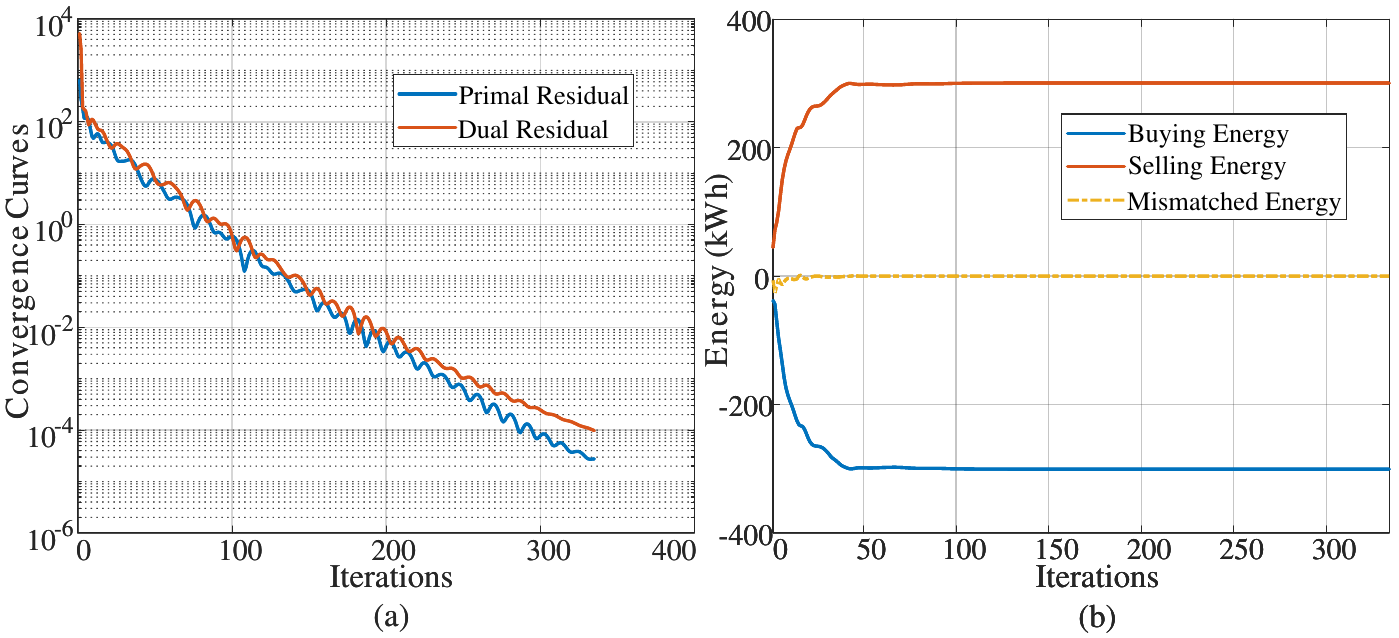}
	\caption{Convergence curves without Byzantine faults: (a) primal and dual residuals, (b) evolution of the traded energy.}
	\label{fig5}
\end{figure}

The effectiveness of the proposed spatial-temporal anomaly detection approach is first verified on the IEEE 15-bus systems with P2P energy trading under different malicious scenarios. As illustrated in Fig. \ref{fig4}, each energy buyer/seller is located at one bus and equipped with an edge computing unit, enabling each prosumer to perform individual computing and to share data with each other. Ideally, when each agent honestly follows the normal distributed optimization protocols and extraneous adversaries do not exist, the normal distributed algorithm will converge to optimal solutions, as illustrated in Fig. \ref{fig5} (a) and Fig. \ref{fig5} (b). As given in Table \ref{tab1}, primal and dual residuals drop to the stopping criteria within 334 iterations, which demonstrates good convergence performance of the normal distributed algorithm. 

\begin{figure}[!hbp]
	\centering
	\includegraphics[width=1.0\linewidth]{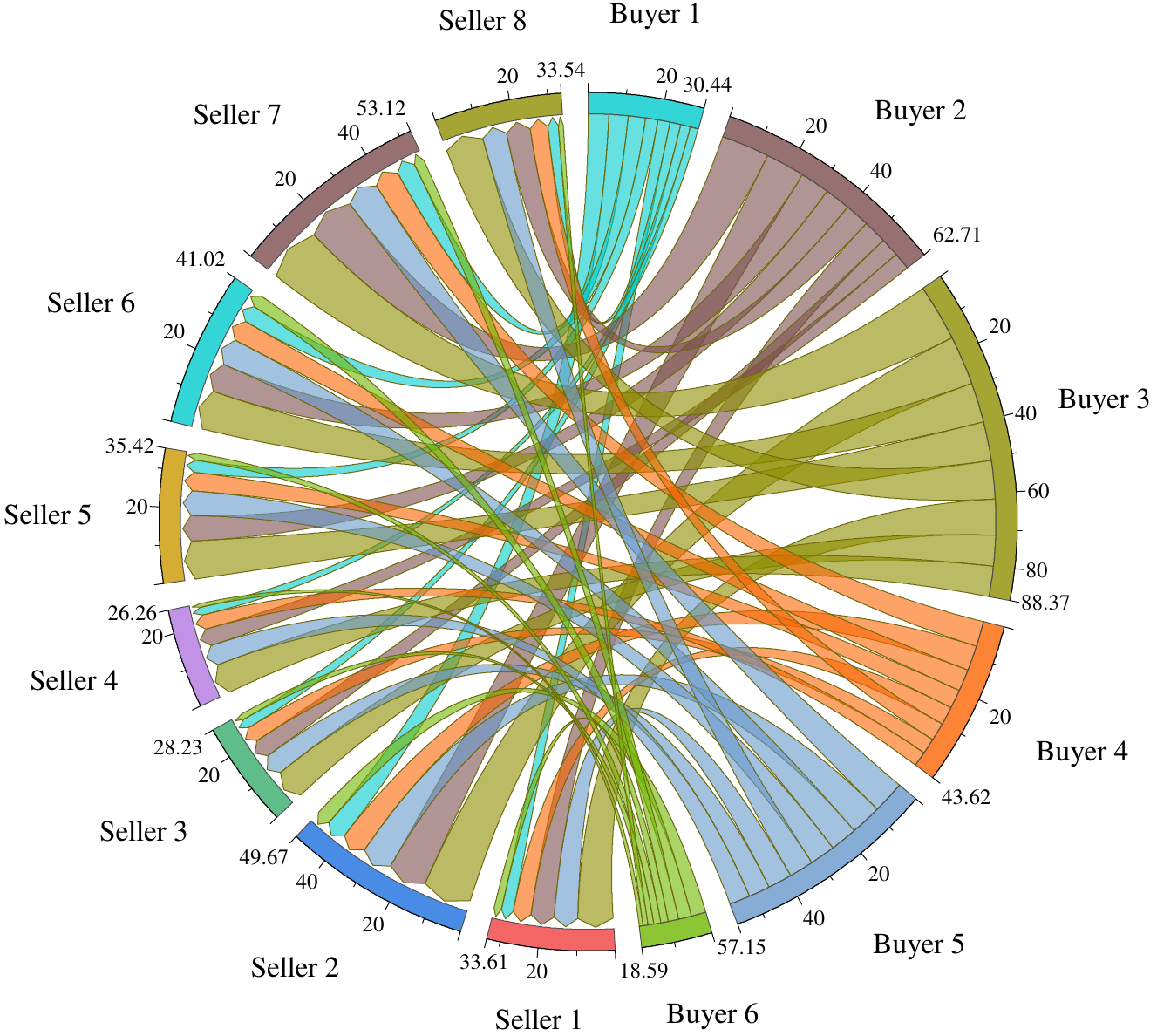}
	\caption{Detailed amounts of traded energy among prosumers.}
	\label{fig6}
\end{figure}

As displayed in Table \ref{tab1}, the amount of traded energy finally evolves to the optimal solution and the mismatched amount of traded energy tends to zero within finite iterations. Specifically, detailed amounts of traded energy for each buyer and seller are exhibited in Fig. \ref{fig6}. Due to distinct parameter settings of the utility functions, reflecting heterogeneous preferences of prosumers, the amounts of traded energy among prosumers vary for different prosumers. For instance, buyer 3 shows a higher inclination to purchase energy from seller 7 compared to buyer 1. However, when the adversary exists in the system, shared data among agents can be manipulated, i.e., the bidding quantity from buyer 3 to seller 7, causing unfair market competitions. Furthermore, severe manipulations via sophisticated false data injections can result in divergence of the normal distributed algorithm and disrupt the normal operation of P2P energy market. To mitigate these malicious impacts, customized anomaly detection mechanisms and countermeasures are pressing. 

\begin{figure}[htbp]
	\centering
	\includegraphics[width=1.0\linewidth]{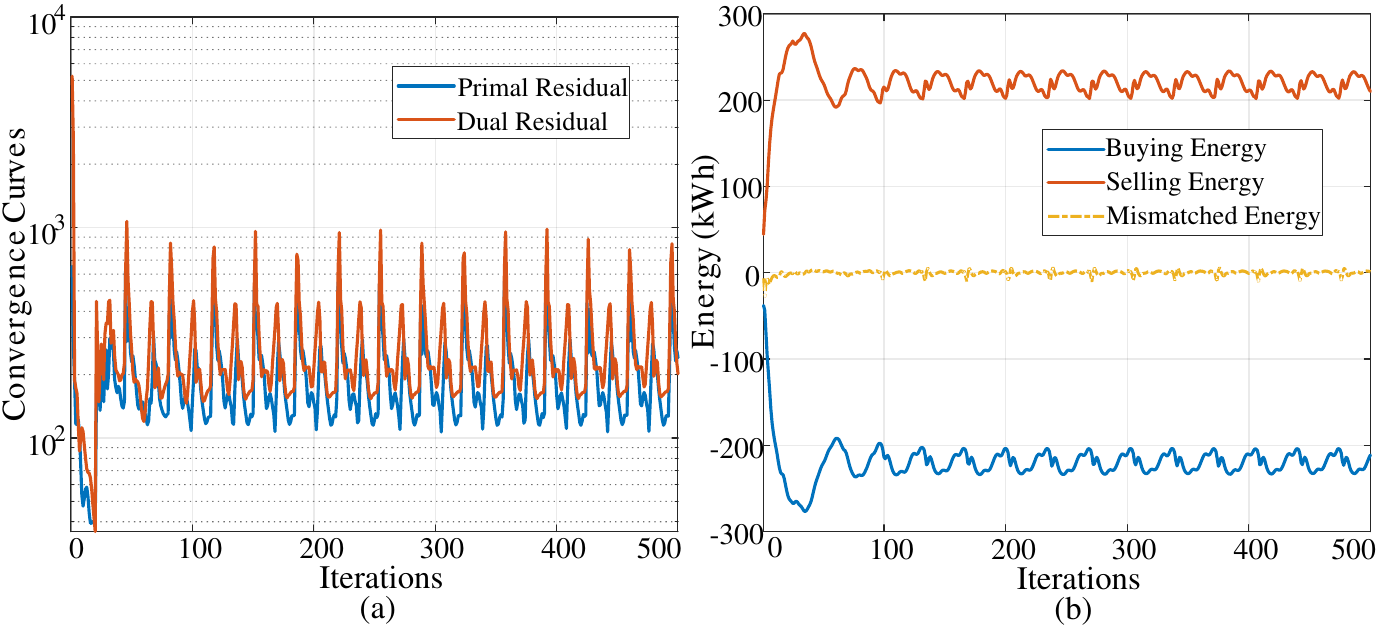}
	\caption{Without the proposed anomaly approach: (a) primal and dual residuals, (b) evolution of the traded energy.}
	\label{fig7}
	\vspace{-0.2cm}
\end{figure}

The proposed online spatial-temporal anomaly detection approach is first verified against dynamic small noise injections. The correlated static false data injection is modeled as below:
\begin{subequations}
	\begin{align}
		&\varkappa = \hat{\varkappa}, \quad \widetilde{\ell}_j=\ell_j+\varkappa \label{fals1}\\
		&\widetilde{P}_j=P_j+R_j\varkappa, \quad \widetilde{Q}_j=Q_j+X_j\varkappa, \label{fals2}
	\end{align}	
\end{subequations}
where $\varkappa$ is the static deviation for $\ell_j$. Considering that the malicious agent manipulates the real-time shared data following \eqref{fals1}-\eqref{fals2} every five iterations. The injected manipulated data, i.e., $\widetilde{\ell}_j$, $\widetilde{P}_j$, and $\widetilde{Q}_j$, will not violate the power flow constraints, which also remain undetectable by the traditional anomaly detection methods \cite{chang2022byzantine,cheng2021resilient}. As illustrated in Fig. \ref{fig7} (a), when $\hat{\varkappa}=200$, the injected correlated false data will cause significant fluctuations of primal and dual residuals, and finally lead to divergence of the distributed P2P energy trading problem. Meanwhile, manipulations of real-time shared power flow data also exert influences on the amount of traded energy as illustrated in Fig. \ref{fig7} (b). As given in Table \ref{tab1}, for the normal case, the optimal total traded amount of energy in the market, i.e., $\Omega$, is about $300.87$kWh and the optimal active power injection at the substation bus, i.e., $p_0$, is $-270.04$kWh. However, when correlated static false data injection attacks happen, the total amount of traded energy oscillates around $200.87$kWh, and active power injection at the substation bus also fluctuates around $-121.19$kWh. Therefore, correlated static false data injection attacks have severe impacts on the normal operation of distributed P2P energy market.
\begin{table}[htbp]
	\centering
	\caption{IEEE 15-bus Systems with Correlated Static False Data Injection}
	\label{tab1}
	\begin{tabular}{ccccc}
		\toprule  
		& $\Omega$ (kWh) &$p_0$ (kWh) & Iteration &Time (s) \\ 
		\cmidrule(r){2-5}
		{Normal Case}&300.88 &-270.04&334 &33.10\\
		{With Attacks}&$\approx200.87$ &$\approx-121.19$&500 &49.30\\		
		{With Detection}&300.88&-270.19&312 &118.54\\
		\bottomrule 
	\end{tabular}
\end{table}\

\begin{figure}[htbp]
	\centering
	\includegraphics[width=1.0\linewidth]{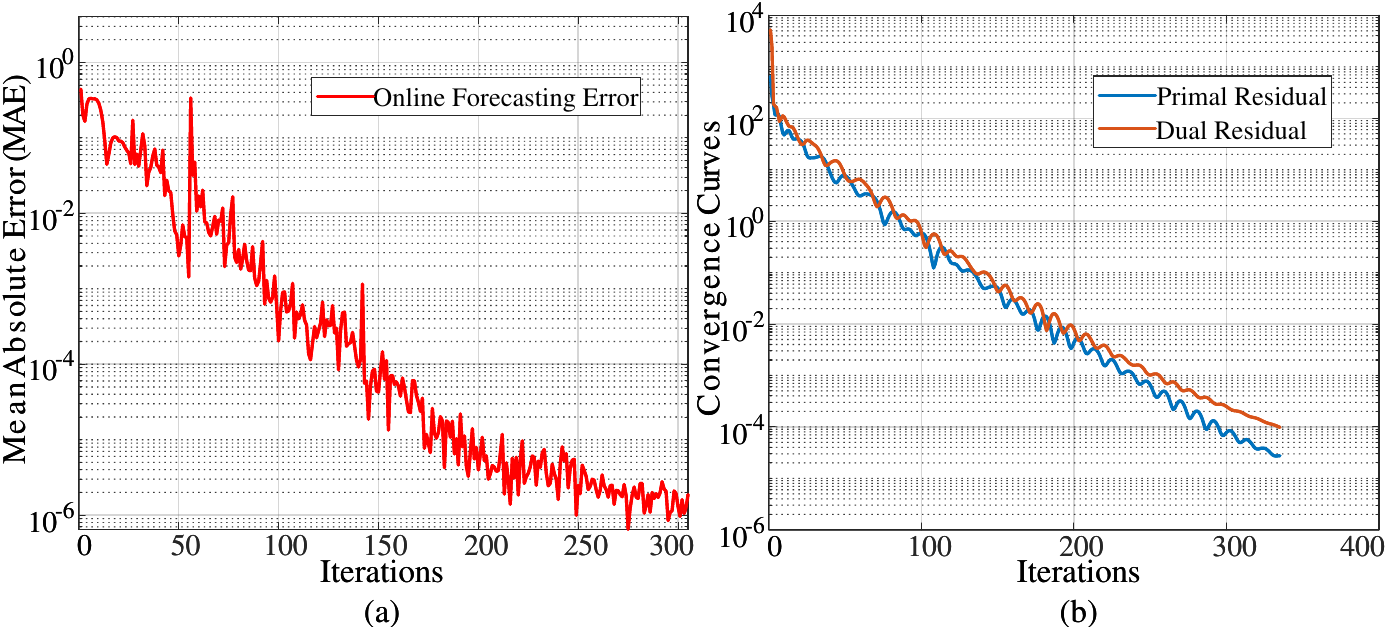}
	\caption{With the proposed anomaly approach: (a) online prediction error, (b) primal and dual residuals.}
	\label{fig8}
\end{figure}

\begin{figure}[htbp]
	\centering
	\includegraphics[width=1.0\linewidth]{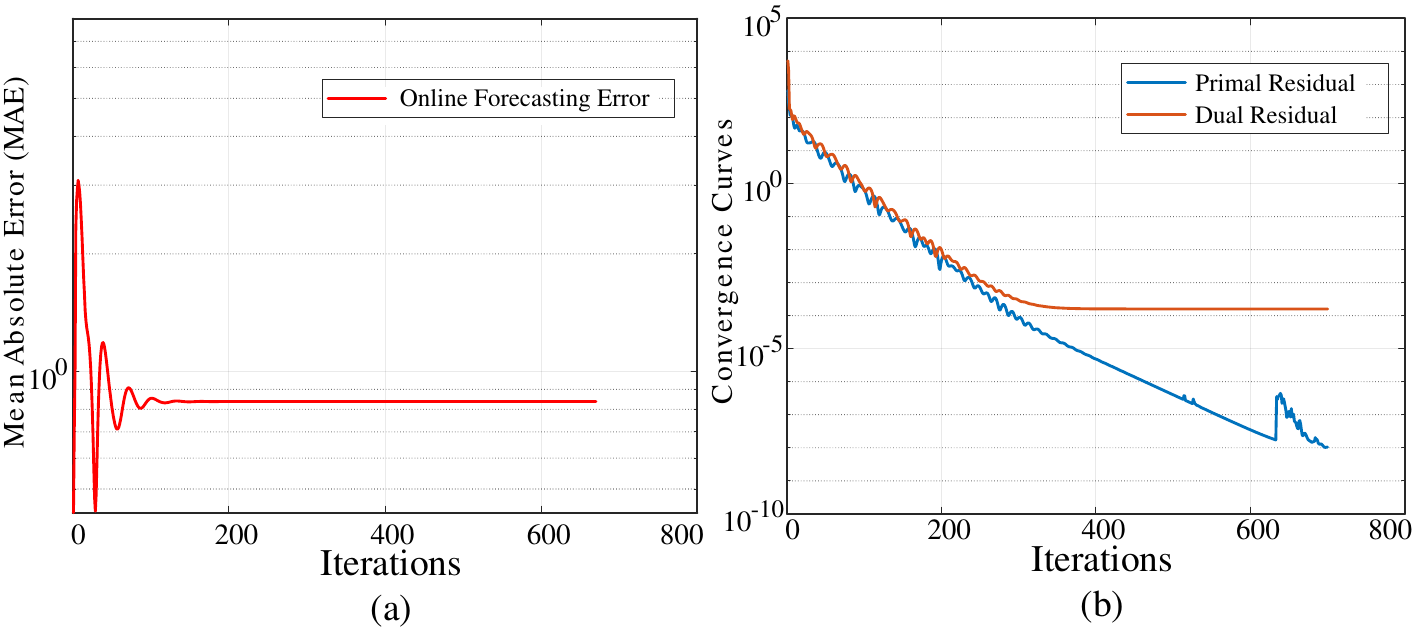}
	\caption{Without the physics-informed mechanism: (a) online prediction error, (b) primal and dual residuals.}
	\label{fig8.1}
\end{figure}

The online spatial-temporal anomaly detection approach is thus proposed to deal with such sophisticated stealthy false data injection attacks. As shown in Fig. \ref{fig8} (a), the mean absolute error for the proposed online detection approach gradually reaches a pretty small value, i.e., $10^{-6}$. The increasing prediction accuracy of the proposed approach is paramount for detecting abnormal data and facilitating convergence of the distributed P2P energy trading problem. As illustrated in Fig. \ref{fig8} (b), primal and dual residuals drop smoothly to the required stopping criteria when the proposed approach is implemented. As given in Table \ref{tab1}, the total amount of traded energy derived by the Byzantine-resilient distributed algorithm safeguarded by the proposed approach approximately equals to the optimal one, so as the active power injection at substation bus. It is found that the Byzantine-resilient distributed algorithm can converge to the required stopping criteria with fewer iterations compared to the normal case. To further validate contributions of the physics-informed design to the overall performance, we conducted additional case studies based on the investigations as in Fig. \ref{fig8}. We keep all the parameter settings the same as the case study as in Fig. \ref{fig8} except for the removal of the physics-informed part in \eqref{op1} for the validation. The numerical results are presented in Fig. \ref{fig8.1}. From these results, we can find that the anomaly detection mechanism without the physics-informed mechanism performs worse as in Fig. \ref{fig8.1} (a), where the prediction accuracy are not comparable with that derived by the proposed approach as in Fig. \ref{fig8.1} (a). The distributed optimization without the physics-informed mechanism can hardly converge to the required stopping criteria, i.e., $\varpi_1 = 10^{-4}$ and $\varpi_2=10^{-4}$, within the given finite 700 iterations as in Fig. \ref{fig8.1} (b). These results validate the necessity of accounting for the spatial correlations in the anomaly detection.

Meanwhile, we concede that the proposed anomaly detection approach requires additional time to update the model parameters and forecast new data. Approximately $3.58X$ more time than that of the normal case is required for the overall Byzantine-resilient distributed algorithm with the anomaly detection approach.

\subsection{Correlated Dynamic Noise Injections}
\begin{figure}[htbp]
	\centering
	\includegraphics[width=1.0\linewidth]{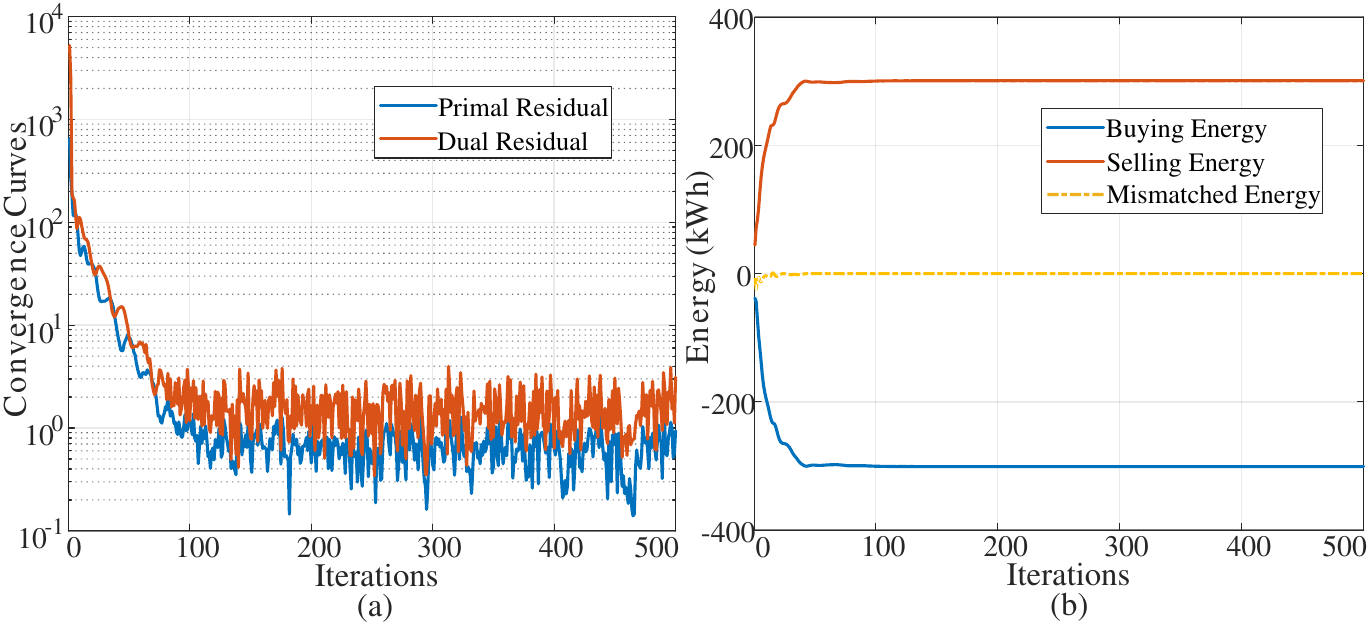}
	\caption{Without the proposed anomaly approach: (a) primal and dual residuals, (b) evolution of the traded energy.}
	\label{fig9}
	\vspace{-0.3cm}
\end{figure}

The effectiveness of the proposed anomaly detection approach is also validated when small dynamic noise injection attacks are present. The static deviation $\varkappa$ in \eqref{fals1} is set to be the random variable as $\varkappa \sim U [0,3]$. The injected false data, i.e., $\widetilde{\ell}_j$, $\widetilde{P}_j$, $\widetilde{Q}_j$, are dynamically correlated and also satisfy the physical network constraints. As shown in Fig. \ref{fig9} (a), the normal distributed algorithm attacked by the dynamic small noises can not converge to required stopping criteria within finite iterations. Due to the redundancy of power system, injected small false data on the power flow has less influence on the clearing process of energy trading. However, the active power injection at the substation bus increases from $-270.04$kWh to about $-266.60$kWh as given in Table \ref{tab2}. Note that if the real-time shared amounts of traded energy from prosumers are manipulated by adversaries, the clearing process of the energy trading can also be substantially influenced. For defending manipulations of these data, it is applicable by adopting a corresponding spatial correlation matrix $\mathcal{W}_i$ required by the proposed online anomaly detection approach. 
\vspace{-0.5cm}

\begin{table}[htbp]
	\centering
	\caption{IEEE 15-bus Systems with Dynamic Small Noise Injections}
	\label{tab2}
	\begin{tabular}{ccccc}
		\toprule  
		&$\Omega$ (kWh) &$p_0$ (kWh) &Iteration & Time (s) \\ 
		\cmidrule(r){2-5}
		{With Attacks}&$\approx300.90$ &$\approx-266.60$&500 &46.83\\		
		{With Detection}&300.88&-270.04&305 &122.62\\
		\bottomrule 
	\end{tabular}
	\vspace{-0.5cm}
\end{table}\

\begin{figure}[htbp]
	\centering
	\includegraphics[width=1.0\linewidth]{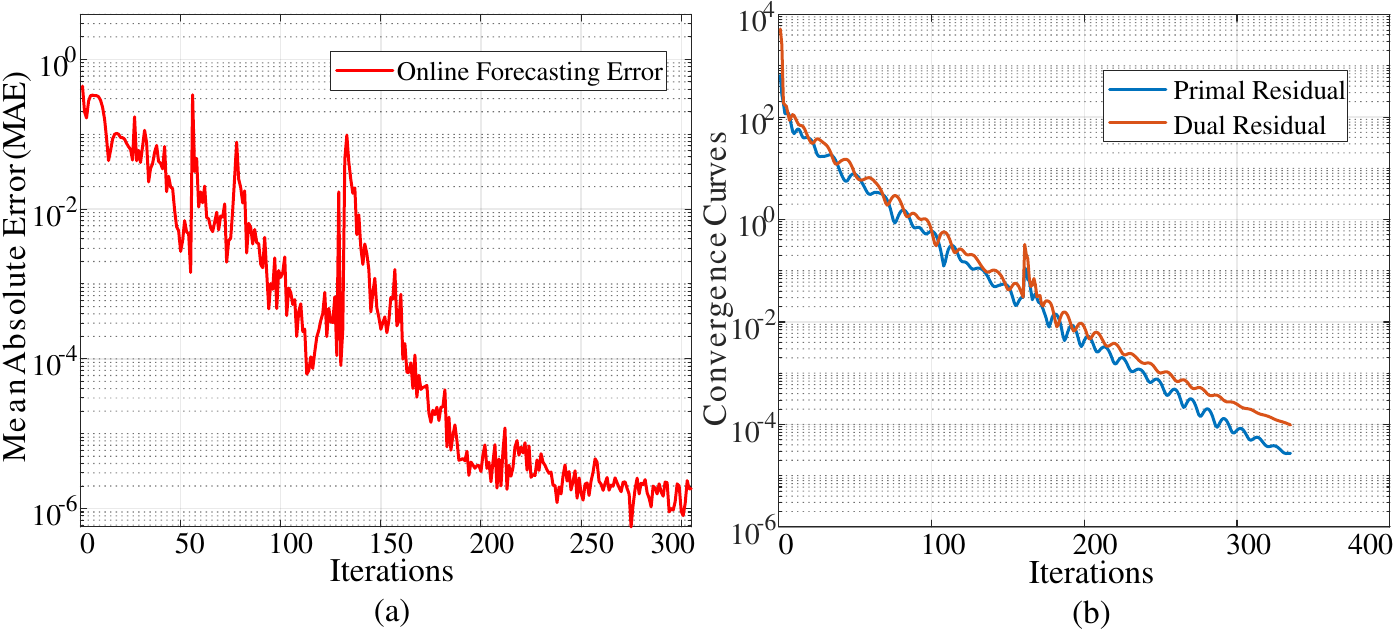}
	\caption{With the proposed anomaly approach: (a) online prediction error, (b) primal and dual residuals.}
	\label{fig10}
	\vspace{-0.6cm}
\end{figure}

As illustrated in Fig. \ref{fig10} (a), the prediction error of the proposed spatial-temporal anomaly detection approach finally drops to around $10^{-6}$, enabling accurate real-time predictions. With the proposed anomaly detection approach, the Byzantine-resilient distributed algorithm converges to the optimal solutions within 305 iterations as shown in Fig. \ref{fig10} (b). As in Table \ref{tab2}, the total amount of traded energy derived by the Byzantine-resilient distributed algorithm equals to the optimal amount, and the active power injection at the substation bus also tends to the optimal one. These results demonstrate that the proposed anomaly detection approach can defend the dynamic small noise injection attacks.

\subsection{Computational Efficiency and Scalability}
\begin{figure}[htbp]
	\centering
	\includegraphics[width=1.0\linewidth]{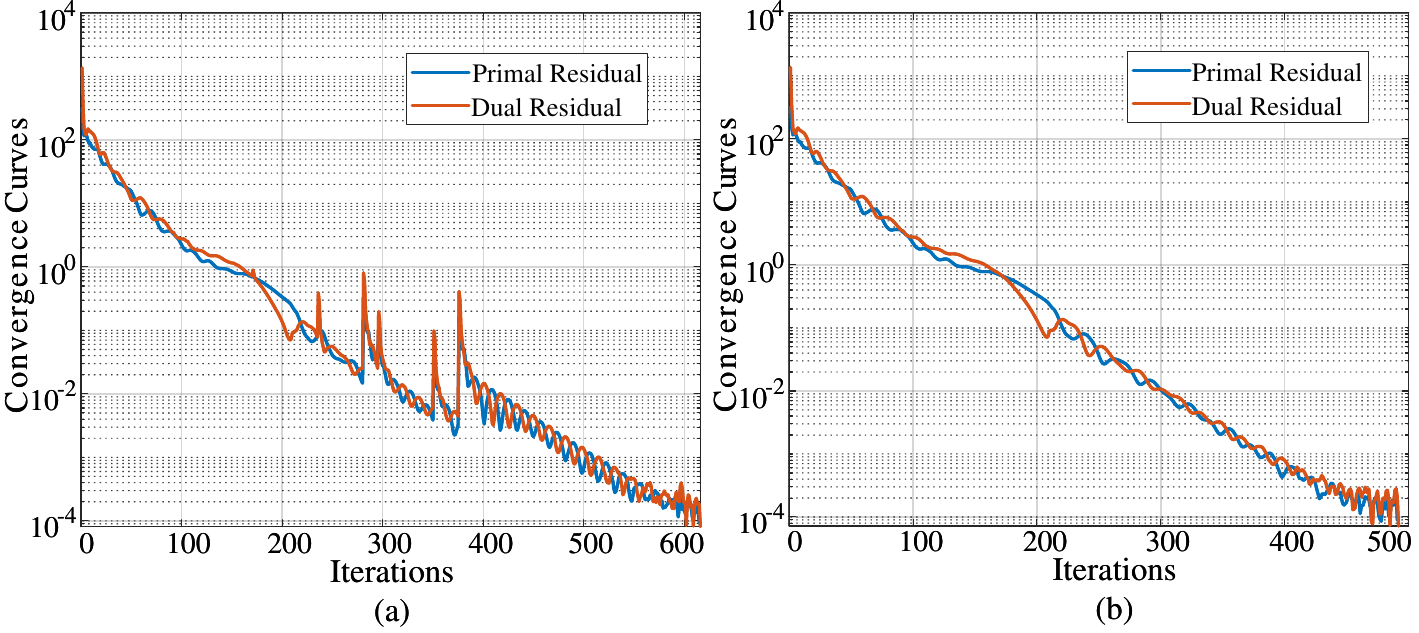}
	\caption{Convergence curves of primal and dual residuals: (a) with dynamic small noise injections, (b) with correlated large static injections.}
	\label{fig11}
	\vspace{-0.3cm}
\end{figure}

The proposed anomaly detection approach is further verified on the larger IEEE 85-bus system with sophisticated false data injections, i.e., the dynamic small noise injection and correlated static false data injection. As illustrated in Figs. \ref{fig11} (a) and (b), the Byzantine-resilient distributed algorithm converges to the required stopping criteria for both cases within finite iterations. As given in Table \ref{tab3}, the total amount of traded energy and active power injection at the substation bus derived by the Byzantine-resilient distributed algorithm are all optimal under two malicious scenarios.

\begin{table}[htbp]
	\centering
	\caption{IEEE 85-bus Systems with False Data Injections}
	\label{tab3}
    \scriptsize
	\begin{tabular}{cccccc}
		\toprule  
		&$\Omega$ (kWh) &$p_0$ (kWh)&Iteration & $T_1$ (s)& $T_2$ (s) \\ 
		\cmidrule(r){2-6}
		{Normal Case}&803.25&5.57 &477 &459.64&5.41\\
		{Small Noise}&803.25&5.57 &616 &776.10&7.75+117.20\\		
		{Large Value}&803.25&5.57 &492 &624.86&6.26+92.40\\
		\bottomrule 
	\end{tabular}
\end{table}

As presented in Table \ref{tab3}, $T_1$ represents the total amount of time for all the agents, where $T_2$ denotes the agent-based average time for each agent when considering the ideal distributed computation. For the dynamic small noise injections, compared to the normal case, additional 139 iterations, i.e., additional total $316.36s$ for all the agents, are needed for the Byzantine-resilient distributed algorithm to reach the required stopping criteria. With correlated static false data injections, the Byzantine-resilient distributed algorithm requires 15 more iterations, i.e., additional $164.36s$ for all the agents, to converge. From the perspective of ideal distributed optimization, the proposed anomaly detection approach incurs an average computation time of $124.95s$ per agent for dynamic small noise injections and $98.66s$ per agent for correlated static false data injections. Notably, the computation time required for the proposed approach remains the same order across different system scales. For instance, the required times for each agent are $92.40s$ for detecting correlated static false data injections and $117.20s$ for detecting dynamic small noise injections in the larger 85-bus system, compared to $85.44s$ and $89.52s$, respectively, in the smaller 15-bus system.

For different system scales, the Byzantine-resilient distributed algorithm with the proposed spatial-temporal anomaly detection approach can derive high-quality solutions within $15$ minutes under different malicious scenarios. Since the proposed anomaly detection approach can be deployed for each prosumer independently, the additional computational time required for the Byzantine-resilient distributed algorithm will not increase exponentially with the system scale. These results demonstrate good scalability of the proposed spatial-temporal anomaly detection approach, making it applicable for the real-time P2P energy trading problem with the clearing intervals of every 15 minute. To further increase the computational efficiency of the distributed computation for individual agents, it is applicable to employ the closed-form solutions for solving sub-problems (6a) to avoid the employment of burdensome commercial solvers \cite{chang2022byzantine,li2018distributed}. Meanwhile, adaptively adjusting the length of online detection window can help boost the computational efficiency for the online anomaly detection mechanism \cite{feng2022fast}.


\section{Conclusions}
To better investigate the impacts of Byzantine faults, we develop a fully distributed P2P energy trading model that accounts for the power losses. For efficiently detecting Byzantine faults and mitigating their impacts, we propose an online spatial-temporal anomaly detection approach by leveraging the domain knowledge-informed online tensor learning approach. We further obtain the closed-form solutions for updating model parameters embedded in the learning method to enhance its computational efficiency. Moreover, we obtain theoretical convergence conditions for the distributed P2P energy trading problem with anomaly detection mechanisms. Numerical simulations demonstrate that the proposed online anomaly detection approach: i) can efficiently detect sophisticated false data injections; ii) ensures convergence, optimality, and good scalability of the distributed P2P energy trading problem.  

Overall, the proposed approach facilitates Byzantine-resilient P2P energy trading by efficiently detecting sophisticated false data injection attacks. Future work aims to further optimize the computational efficiency and performance of the proposed P2P energy trading approach by developing closed-form solutions for sub-problems, designing adaptive methods to dynamically adjust the window length for the online anomaly detection, and exploring systematic approaches for optimally deciding other hyper-parameters.
 
{\appendix
	
     \subsection{Closed-Form Solutions for Online Tensor Learning}
    	Solving ${\cal P}_6$ is computationally expensive. By exploring the special structure of this optimization problem, we can derive closed-form solutions for updating $\Delta^d\mathcal{\hat{G}}_{i,t}^{(n)}$ and $\mathbb{\hat{\epsilon}}_{i,t-m}^{(n)}$ via the first-order partial derivatives of \eqref{op1}-\eqref{op2}:
    	
    	\begin{align}
    		\Delta^d\mathcal{\hat{G}}_{i,t}^{(n)} = & \frac{1}{\hat{\mathbf{A}}_i^{(-n)}\mathcal{W}_i\mathcal{W}_i^T\hat{\mathbf{A}}_i^{(-n)^T}+2I} (\hat{\mathbf{A}}_i^{(n)^T}\Delta^d\mathcal{\hat{D}}_{i,t}^{(n)}\hat{\mathbf{A}}_i^{(-n)^T} \notag \\
    		&+\sum_{m=1}^p\psi_m \Delta^d\mathcal{\hat{G}}_{i,t-m}^{(n)} -\sum_{m=1}^q \theta_m \mathbb{\hat{\epsilon}}_{i,t-m}^{(n)}) \label{up1}
    	\end{align}
    	\begin{align}
    		\mathbb{\hat{\epsilon}}_{i,t-m}^{(n)} = &\frac{1}{(p+q+d+1-\hat{k}_n)\theta_m}(\sum\limits_{\substack{t=p+d\\+q+1}}^{\hat{k}_n}(\Delta^d\mathcal{\hat{G}}_{i,t}^{(n)} \notag \\
    		&-\sum_{m=1}^p \psi_m \Delta^d\mathcal{\hat{G}}_{i,t-m}^{(n)} + \sum_{m=1}^q \theta_m \mathbb{\hat{\epsilon}}_{i,t-m}^{(n)})). \label{up2}
    	\end{align}
    	For the updates of $\hat{\mathbf{A}}_i^{(n)}$, the optimization sub-problem is formulated as follows:
    	\begin{subequations}
    		\begin{align}
    			\mathcal{P}_7: \quad \min\limits_{\hat{\mathbf{A}}_i^{(n)}} &\sum\limits_{\substack{t=p+d\\+q+1}}^{\hat{k}_n}\sum_{n=1}^{N} (\frac{1}{2} ||\Delta^d\mathcal{\hat{G}}_{i,t}^{(n)}-\hat{\mathbf{A}}_i^{(n)^T}\Delta^d\mathcal{\hat{D}}_{i,t}^{(n)}\hat{\mathbf{A}}_i^{(-n)^T}||_F^2 ) \label{ts1} \\
    			\mathbf{s.t.} &\quad  \hat{\mathbf{A}}_i^{(n)^T}\hat{\mathbf{A}}_i^{(n)}=\mathbf{I}, n=1,...,N. \label{ts2} 
    		\end{align}
    	\end{subequations}
    	The closed-form solutions for $\hat{\mathbf{A}}_i^{(n)}$ can be obtained through an equivalent optimization problem:
    	\begin{align}
    		\resizebox{0.88\hsize}{!}{$ \hat{\mathbf{A}}_i^{(n)} = \mathop{\text{argmax}}\limits_{\hat{\mathbf{A}}_i^{(n)}} \sum\limits_{\substack{t=p+d\\+q+1}}^{\hat{k}_n} trace (\hat{\mathbf{A}}_i^{(n)^T}\Delta^d\mathcal{\hat{D}}_{i,t}^{(n)}\hat{\mathbf{A}}_i^{(-n)^T}\Delta^d\mathcal{\hat{G}}_{i,t}^{(n)}). $}
    	\end{align}
    	Thus we can obtain closed-form solutions for $\hat{\mathbf{A}}_i^{(n)}$ as below:
    	\begin{align}
    		\hat{\mathbf{A}}_i^{(n)} = \hat{\mathbf{U}}_i^{(n)}\hat{\mathbf{V}}_i^{(n)}, \label{up3}
    	\end{align}
    	where $\hat{\mathbf{U}}_i^{(n)}$ and $\hat{\mathbf{V}}_i^{(n)}$ are the left and right vectors of  the singular value decomposition (SVD) of $\Delta^d\mathcal{\hat{D}}_{i,t}^{(n)}\hat{\mathbf{A}}_i^{(-n)^T}\Delta^d\mathcal{\hat{G}}_{i,t}^{(n)}$. After obtaining closed-form solutions of $\Delta^d\mathcal{\hat{G}}_{i,t}^{(n)}$, $\mathbb{\hat{\epsilon}}_{i,t-m}^{(n)}$, and $\hat{\mathbf{A}}_i^{(n)}$, parameters $\psi_m$ and $\theta_m$ can be estimated through the Yule-Walker method \cite{friedlander1984modified}.

	\subsection{Proof of Theorem 1} 
	The difference of the Lagrangian function for all the prosumers over two consecutive iterations can be denoted as:

\begin{align}
      &\sum\limits_{i \in {\cal N}_p}\mathcal{L}_i(\mathbf{x}_i^{k+1}, \mathbf{y}_{i(j)}^{k+1},\mathbf{\mu}_{i(j)}^{k+1})-\sum\limits_{i \in {\cal N}_p}\mathcal{L}_i(\mathbf{x}_i^{k}, \mathbf{y}_{i(j)}^{k},\mathbf{\mu}_{i(j)}^{k}) \notag\\
      = &\sum\limits_{i \in {\cal N}_p}\mathcal{L}_i(\mathbf{x}_i^{k+1}, \mathbf{y}_{i(j)}^{k+1},\mathbf{\mu}_{i(j)}^{k+1})-\sum\limits_{i \in {\cal N}_p}\mathcal{L}_i(\mathbf{x}_i^{k+1}, \mathbf{y}_{i(j)}^{k+1},\mathbf{\mu}_{i(j)}^{k}) \notag\\
      &+\sum\limits_{i \in {\cal N}_p}\mathcal{L}_i(\mathbf{x}_i^{k+1}, \mathbf{y}_{i(j)}^{k+1},\mathbf{\mu}_{i(j)}^{k})-\sum\limits_{i \in {\cal N}_p}\mathcal{L}_i(\mathbf{x}_i^{k+1}, \mathbf{y}_{i(j)}^{k},\mathbf{\mu}_{i(j)}^{k}) \notag\\
      &+\sum\limits_{i \in {\cal N}_p}\mathcal{L}_i(\mathbf{x}_i^{k+1}, \mathbf{y}_{i(j)}^{k},\mathbf{\mu}_{i(j)}^{k})-\sum\limits_{i \in {\cal N}_p}\mathcal{L}_i(\mathbf{x}_i^{k}, \mathbf{y}_{i(j)}^{k},\mathbf{\mu}_{i(j)}^{k}).  \label{pf1}
\end{align}	

For different terms in the above relation, they hold that:
\begin{align}
	&\sum\limits_{i \in {\cal N}_p}\mathcal{L}_i(\mathbf{x}_i^{k+1}, \mathbf{y}_{i(j)}^{k+1},\mathbf{\mu}_{i(j)}^{k+1})-\sum\limits_{i \in {\cal N}_p}\mathcal{L}_i(\mathbf{x}_i^{k+1}, \mathbf{y}_{i(j)}^{k+1},\mathbf{\mu}_{i(j)}^{k})  \notag\\
	&\le\frac{2}{\eta_i}\sum\limits_{i \in {\cal N}_p}\sum_{j \in \mathcal{N}_i}  \mathcal{Q}_i^{-2}(\lambda_i^2||\mathbf{y}_{i(j)}^{k+1}-\mathbf{y}_{i(j)}^{k}||_2^2) \notag\\
    &+2\eta_i\sum\limits_{i \in {\cal N}_p}(\lambda_i^2||\mathbf{x}_i^{k+1}-\mathbf{x}_i^{k}||_2^2) \label{pf2}
\end{align}	
\vspace{-0.5cm}
\begin{align}
	&\sum\limits_{i \in {\cal N}_p}\mathcal{L}_i(\mathbf{x}_i^{k+1}, \mathbf{y}_{i(j)}^{k},\mathbf{\mu}_{i(j)}^{k})-\sum\limits_{i \in {\cal N}_p}\mathcal{L}_i(\mathbf{x}_i^{k}, \mathbf{y}_{i(j)}^{k},\mathbf{\mu}_{i(j)}^{k})  \notag\\
	&\le - \frac{\Upsilon (\delta_i+\eta_i)}{2} \sum\limits_{i \in {\cal N}_p}||\mathbf{x}_i^{k+1}-\mathbf{x}_i^{k}||_2^2 \label{pf3}
\end{align}
\vspace{-0.4cm}
\begin{align}
	&\sum\limits_{i \in {\cal N}_p}\mathcal{L}_i(\mathbf{x}_i^{k+1}, \mathbf{y}_{i(j)}^{k+1},\mathbf{\mu}_{i(j)}^{k})-\sum\limits_{i \in {\cal N}_p}\mathcal{L}_i(\mathbf{x}_i^{k+1}, \mathbf{y}_{i(j)}^{k},\mathbf{\mu}_{i(j)}^{k})  \notag\\
	&\le - \frac{\Upsilon \eta_i}{2}\sum\limits_{i \in {\cal N}_p}\sum_{j \in \mathcal{N}_i}||\mathbf{y}_{i(j)}^{k+1}-\mathbf{y}_{i(j)}^{k}||_2^2. \label{pf4}
\end{align}	
The optimization problem in \eqref{sb1} can be reformulated as:
\begin{align}
\mathbf{x}_i^{k+1}&=\mathop{\text{arg min}}\limits_{\mathbf{x}_i \in \mathcal{X}_i} \mathcal{L}_i(\mathbf{x}_i, \mathbf{y}_{i(j)}^{k},\mathbf{\mu}_{i(j)}^{k}) \notag\\
&=\mathop{\text{arg min}}\limits_{\mathbf{x}_i} \mathcal{L}_i(\mathbf{x}_i, \mathbf{y}_{i(j)}^{k},\mathbf{\mu}_{i(j)}^{k}) + \mathbb{I}_{\mathcal{X}_i} (\mathbf{x}_i). \label{pf5}
\end{align}	
The first-order optimality condition for \eqref{pf5} can be written as:
\begin{align}
 	0 \in \partial_{\mathbf{x}_i} (\mathcal{L}_i(\mathbf{x}_i^k, \mathbf{y}_{i(j)}^{k},\mathbf{\mu}_{i(j)}^{k}) +  \mathbb{I}_{\mathcal{X}_i} (\mathbf{x}_i^k)), \label{pf6}
\end{align}
where $\partial(\centerdot)$ is the subgradient operator. $\mathcal{L}_i(\mathbf{x}_i, \mathbf{y}_{i(j)}^{k},\mathbf{\mu}_{i(j)}^{k})+ \mathbb{I}_{\mathcal{X}_i} (\mathbf{x}_i)$ is $(\delta_i+\eta_i)$-strongly convex, and it holds that:
\begin{align}
	\mathcal{L}_i&(\mathbf{x}_i^{k+1}, \mathbf{y}_{i(j)}^{k},\mathbf{\mu}_{i(j)}^{k}) +  \mathbb{I}_{\mathcal{X}_i} (\mathbf{x}_i^{k+1}) \notag \\
	&\le \mathcal{L}_i(\mathbf{x}_i^{k}, \mathbf{y}_{i(j)}^{k},\mathbf{\mu}_{i(j)}^{k}) +  \mathbb{I}_{\mathcal{X}_i} (\mathbf{x}_i^{k})- \frac{\eta_i+\delta_i}{2}||\mathbf{x}_i^{k+1}-\mathbf{x}_i^{k}||_2^2 \notag \\
	&+<\partial(\mathcal{L}_i(\mathbf{x}_i^{k+1}, \mathbf{y}_{i(j)}^{k},\mathbf{\mu}_{i(j)}^{k})),\mathbf{x}_i^{k+1}-\mathbf{x}_i^{k}>,
\end{align}
when $||d_{i,k+1}^{p}-d_{i,k}||_2\le \lambda_i||d_{i,k}-d_{i,k-1}||_2 $ or $||d_{i,k+1}^{r}-d_{i,k+1}^{p}||_2 \le \varphi_i$, 
$d_{i,k+1}$ is either the true or predicted value, which means that $d_{i,k+1} \in \mathcal{X}_i$. It holds that $\mathbb{I}_{\mathcal{X}_i} (\mathbf{x}_i^{k+1})=\mathbb{I}_{\mathcal{X}_i} (\mathbf{x}_i^{k}) =0 $ for all updates, and It can be obtained that: 
\begin{align}
	&\sum\limits_{i \in {\cal N}_p} \mathcal{L}_i(\mathbf{x}_i^{k+1}, \mathbf{y}_{i(j)}^{k},\mathbf{\mu}_{i(j)}^{k})-\sum\limits_{i \in {\cal N}_p} \mathcal{L}_i(\mathbf{x}_i^{k}, \mathbf{y}_{i(j)}^{k},\mathbf{\mu}_{i(j)}^{k})  \notag\\
	&\le - \frac{\Upsilon (\delta_i+\eta_i)}{2} \sum\limits_{i \in {\cal N}_p} \sum_{k=0}^{K}||\mathbf{x}_i^{k+1}-\mathbf{x}_i^{k}||_2^2.
\end{align}
This completes the proof for \eqref{pf3}. Meanwhile, \eqref{pf4} can be obtained following the similar reasoning. For relation \eqref{pf2}, it can be obtained as follows:
\begin{align}
&\sum\limits_{i \in {\cal N}_p}\mathcal{L}_i(\mathbf{x}_i^{k+1}, \mathbf{y}_{i(j)}^{k+1},\mathbf{\mu}_{i(j)}^{k+1})-\sum\limits_{i \in {\cal N}_p}\mathcal{L}_i(\mathbf{x}_i^{k+1}, \mathbf{y}_{i(j)}^{k+1},\mathbf{\mu}_{i(j)}^{k}) \notag \\
&=\sum\limits_{i \in {\cal N}_p}\sum_{j \in \mathcal{N}_i} <\mathbf{\mu}_{i(j)}^{k+1}-\mathbf{\mu}_{i(j)}^{k},\mathbf{x}_i^{k+1}-\mathbf{y}_{i(j)}^{k+1}> \notag \\
&=\sum\limits_{i \in {\cal N}_p}\sum_{j \in \mathcal{N}_i}\frac{1}{\eta_i}||\mathbf{\mu}_{i(j)}^{k+1}-\mathbf{\mu}_{i(j)}^{k}||_2^2 \notag \\
&\mathop{=}\limits^{1}\sum\limits_{i \in {\cal N}_p}\sum_{j \in \mathcal{N}_i}\frac{1}{\eta_i}||\mathcal{Q}_i^{-1}\mathbf{y}_{i(j)}^{k+2}-\mathcal{Q}_i^{-1}\mathbf{y}_{i(j)}^{k+1}+\eta_i\mathbf{x}_i^{k+2}-\eta_i\mathbf{x}_i^{k+1}||_2^2 \notag\\
&\mathop{\le}\limits^{2} \sum\limits_{i \in {\cal N}_p}\sum_{j \in \mathcal{N}_i}\frac{2}{\eta_i}(\mathcal{Q}_i^{-2}||\mathbf{y}_{i(j)}^{k+2}-\mathbf{y}_{i(j)}^{k+1}||_2^2+\eta_i^2||\mathbf{x}_i^{k+2}-\mathbf{x}_i^{k+1}||_2^2) \notag\\
&\mathop{\le}\limits^{3} \sum\limits_{i \in {\cal N}_p}\sum_{j \in \mathcal{N}_i}\frac{2}{\eta_i}\mathcal{Q}_i^{-2}||\lambda_i(\mathbf{y}_{i(j)}^{k+1}-\mathbf{y}_{i(j)}^{k})||_2^2 \notag\\
&\quad +\sum\limits_{i \in {\cal N}_p}2\eta_i||\lambda_i(\mathbf{x}_i^{k+1}-\mathbf{x}_i^{k})||_2^2, \label{spf1}
\end{align}	
where the first marked relation is derived from the update formula $\mathbf{\mu}_{i(j)}^{k}=-\mathcal{Q}_i^{-1}\mathbf{y}_{i(j)}^{k+1}-\eta_i\mathbf{x}_i^{k+1}$. The third relation is derived by considering the update rule \eqref{ref5} and \eqref{ref6} for the anomaly data. The Young's inequality \cite{ando1995matrix} is leveraged, i.e.,
\begin{align}
  \mathbf{a}^T\mathbf{b} \le \frac{1}{2\xi}||\mathbf{a}||_2^2+\frac{\xi}{2}||\mathbf{b}||_2^2,
\end{align}
which holds true for any $\mathbf{a}$, $\mathbf{b}$, and $\xi$. Let $\xi=1$ in the Young's inequality, allied with the triangle inequality, the second marked relation can be derived. 

To obtain the final result, we can take a sum of \eqref{pf1} from iteration 1 to $K$.
\begin{align}
&\sum\limits_{i \in {\cal N}_p}\mathcal{L}_i(\mathbf{x}_i^{K}, \mathbf{y}_{i(j)}^{K},\mathbf{\mu}_{i(j)}^{K})-\sum\limits_{i \in {\cal N}_p}\mathcal{L}_i(\mathbf{x}_i^{1}, \mathbf{y}_{i(j)}^{1},\mathbf{\mu}_{i(j)}^{1}) \notag\\
& \le \sum_{k=1}^{K}\sum\limits_{i \in {\cal N}_p}\sum_{j \in \mathcal{N}_i} (\frac{2\mathcal{Q}_i^{-2}\lambda_i^2}{\eta_i} -\frac{\Upsilon\eta_i}{2})||\mathbf{y}_{i(j)}^{k+1}-\mathbf{y}_{i(j)}^{k}||_2^2 \notag\\
& +\sum_{k=1}^{K}\sum\limits_{i \in {\cal N}_p} (2\eta_i\lambda_i^2-\frac{\Upsilon(\delta_i+\eta_i)}{2})||\mathbf{x}_i^{k+1}-\mathbf{x}_i^{k}||_2^2.  \label{pf7}
\end{align}

For the Lagrangian function, it can be lower bounded:
\begin{align}
	&\sum\limits_{i \in {\cal N}_p}\mathcal{L}_i(\mathbf{x}_i^{K}, \mathbf{y}_{i(j)}^{K},\mathbf{\mu}_{i(j)}^{K})=\sum\limits_{i \in {\cal N}_p}\{f_i(\mathbf{x}_i^{K}) \notag\\ 
	&\quad+\sum\limits_{j \in {\cal N}_i}<\mathbf{\mu}_{i(j)}^{K}, \mathbf{x}_i^{K}-\mathbf{y}_{i(j)}^{K}>+\frac{\eta_i}{2}||\mathbf{x}_i^{K}-\mathbf{y}_{i(j)}^{K}||_2^2\} \notag\\
	&=\sum\limits_{i \in {\cal N}_p}\{f_i(\mathbf{x}_i^{K})-\sum\limits_{j \in {\cal N}_i}<\mathcal{Q}_i^{-1}\mathbf{y}_{i(j)}^{K+1}+\eta_i\mathbf{x}_i^{K+1}, \mathbf{x}_i^{K}-\mathbf{y}_{i(j)}^{K}> \notag\\ &\quad+\frac{\eta_i}{2}||\mathbf{x}_i^{K}-\mathbf{y}_{i(j)}^{K}||_2^2\} \notag\\
	&\mathop{\geq} \sum\limits_{i \in {\cal N}_p}\{f_i(\mathbf{x}_i^{K})+\sum\limits_{j \in {\cal N}_i} -\frac{1}{2}||\mathcal{Q}_i^{-1}\mathbf{y}_{i(j)}^{K+1}+\eta_i\mathbf{x}_i^{K+1}||_2^2 \notag\\ 
	&\quad+\frac{\eta_i-1}{2}||\mathbf{x}_i^{K}-\mathbf{y}_{i(j)}^{K}||_2^2\} \notag\\
	&=\Theta_1-\sum\limits_{i \in {\cal N}_p}\sum\limits_{j \in {\cal N}_i}\frac{1}{2}||\mathcal{Q}_i^{-1}\mathbf{y}_{i(j)}^{K+1}+\eta_i\mathbf{x}_i^{K+1}||_2^2 \notag\\
	& \mathop{\geq}\Theta_1-\sum\limits_{i \in {\cal N}_p}\sum\limits_{j \in {\cal N}_i}\frac{\Theta_2}{2}\text{diam}^2(\mathcal{X}_i) > -\infty,  \label{pf8}
\end{align}
where $\Theta_1$ and $\Theta_2$ are constants. Then the Lagrangian function $\sum\limits_{i \in {\cal N}_p}\mathcal{L}_i(\mathbf{x}_i^{K}, \mathbf{y}_{i(j)}^{K},\mathbf{\mu}_{i(j)}^{K})$ is lower bounded when $K \rightarrow \infty$.

Substituting \eqref{pf8} into \eqref{pf7}, we can obtain the relations:
\begin{align}
     &\sum_{k=1}^{K}\sum\limits_{i \in {\cal N}_p}\sum_{j \in \mathcal{N}_i} (\frac{\Upsilon\eta_i}{2}-\frac{2\mathcal{Q}_i^{-2}\lambda_i^2}{\eta_i})||\mathbf{y}_{i(j)}^{k+1}-\mathbf{y}_{i(j)}^{k}||_2^2 \notag\\
	& +\sum_{k=1}^{K}\sum\limits_{i \in {\cal N}_p} (\frac{\Upsilon(\delta_i+\eta_i)}{2}-2\eta_i\lambda_i^2)||\mathbf{x}_i^{k+1}-\mathbf{x}_i^{k}||_2^2 \notag\\ 
	& \le \sum\limits_{i \in {\cal N}_p}\mathcal{L}_i(\mathbf{x}_i^{1}, \mathbf{y}_{i(j)}^{1},\mathbf{\mu}_{i(j)}^{1})-\sum\limits_{i \in {\cal N}_p}\mathcal{L}_i(\mathbf{x}_i^{K}, \mathbf{y}_{i(j)}^{K},\mathbf{\mu}_{i(j)}^{K}) \notag \\
	&\le \infty,  \label{pf9}
\end{align}
where the left hand side (LHS) of \eqref{pf9} is positive and increase with $k$ under the conditions \eqref{cond1}. Since the right hand side (RHS) of \eqref{pf9} is finite, we can draw the conclusions:
\begin{subequations}
\begin{align}
	&\mathop{\text{lim}}_{k \rightarrow \infty} ||\mathbf{x}_i^{k+1}-\mathbf{x}_i^{k}||_2^2 \rightarrow 0, i \in \mathcal{N}_p  \\
	&\mathop{\text{lim}}_{k \rightarrow \infty} ||\mathbf{y}_{i(j)}^{k+1}-\mathbf{y}_{i(j)}^{k}||_2^2 \rightarrow 0, i \in \mathcal{N}_p, j \in \mathcal{N}_i.
\end{align}
Applying the relation \eqref{spf1}, we can obtain:
\begin{align}
	&\mathop{\text{lim}}_{k \rightarrow \infty} ||\mathbf{\mu}_{i(j)}^{k+1}-\mathbf{\mu}_{i(j)}^{k}||_2^2 \rightarrow 0, i \in \mathcal{N}_p, j \in \mathcal{N}_i \\
	&\mathop{\text{lim}}_{k \rightarrow \infty} ||\mathbf{x}_i^{k}-\mathbf{y}_{i(j)}^{k}||_2^2 \rightarrow 0, i \in \mathcal{N}_p, j \in \mathcal{N}_i.
\end{align}
\end{subequations}

Let $\mathbf{x}_i^{*}$, $\mathbf{y}_{i(j)}^{*}$, and $\mathbf{\mu}_{i(j)}^{*}$ denote the limit points of the sequences $\{\mathbf{x}_i^{k}\}$, $\{\mathbf{y}_{i(j)}^{k}\}$, $\{\mathbf{\mu}_{i(j)}^{k}\}$. Then  $\mathbf{x}_i^{*}$, $\mathbf{y}_{i(j)}^{*}$, and $\mathbf{\mu}_{i(j)}^{*}$ is the stationary solution of problem \cite{chang2016asynchronous} and satisfy:
\begin{subequations}
\begin{align}
	&\resizebox{0.85\hsize}{!}{$0 \in \partial_{\mathbf{x}_i} (\mathcal{L}_i(\mathbf{x}_i^*, \mathbf{y}_{i(j)}^{*},\mathbf{\mu}_{i(j)}^{*}) + \mathbb{I}_{\mathcal{X}_i} (\mathbf{x}_i^*)), i \in \mathcal{N}_p, j \in \mathcal{N}_i  $}\\
    &\mathbf{\mu}_{i(j)}^{*}=-\mathcal{Q}_i^{-1}\mathbf{y}_{i(j)}^{*}-\eta_i\mathbf{x}_i^{*}, i \in \mathcal{N}_p, j \in \mathcal{N}_i \\  
    &\mathbf{x}_i^*=\mathbf{y}_{i(j)}^{*} \in \mathcal{X}_i, i \in \mathcal{N}_p, j \in \mathcal{N}_i .
\end{align}
\end{subequations} 
}

\bibliographystyle{IEEEtran}
\bibliography{index}

\end{document}